\documentclass{iopjournal}
\usepackage{natbib}

\begin{document}

\articletype{Topical Review} 

\title{Core Collapse Supernova Modeling: The Next Ten Years}

\author{Anthony Mezzacappa$^1$
}

\affil{$^1$Department of Physics and Astronomy, University of Tennessee, Knoxville, USA}

\email{mezz@utk.edu}


\begin{abstract}
Core collapse supernova modeling has advanced considerably since the first numerical simulations were performed sixty years ago. In particular, the last decade has brought us sophisticated three-dimensional models with significant predictive capabilities---e.g., for core collapse supernova gravitational wave emission. The six decades of modeling have shown us the importance of individual components of these general relativistic neutrino radiation magnetohydrodynamics events---specifically, the importance of neutrino kinetics, fluid instabilities, magnetic fields, strong gravity, and the nuclear equation of state and neutrino--matter interactions calculated in a manner consistent with the equation of state. They have also shown us that simulation outcomes are sensitive to variations in the treatment of these ingredients, demanding a level of rigor that has not yet been consistently met by modelers. The efficacy of the neutrino shock reheating mechanism for core collapse supernovae has been demonstrated. The models now require an improved quantitative predictive capability, which will be achieved through increased sophistication in the treatment of model components, both macroscopic (e.g., strong-field gravity) and microscopic (e.g., neutrino--matter interactions). Advancement of core collapse supernova theory will also require the cooperation of modelers in other fields, especially stellar evolution and nuclear theory, to meet the level of rigor required to make the most of the eventuality of a Galactic core collapse supernova and its multimessenger emissions.
\end{abstract}

\section{Progress to Date}

The first numerical simulations of core collapse supernovae were conducted sixty years ago by \citet{CoWh66}. The first fifty of those sixty years seemed like an eternity of failure in an attempt to identify the mechanism driving these stellar explosions. However, the last decade has brought great success. Three-dimensional models have achieved a certain level of sophistication during this time. With that, a consensus has developed across core collapse supernova modeling groups that core collapse supernovae can be driven by neutrinos. More specifically, that neutrino shock reheating as proposed originally by \citet{Wilson1985}, aided by proto-neutron star convection, turbulent neutrino-driven convection, the standing accretion shock instability, rotation, and magnetic fields, could drive these explosions in progenitors varying in mass, metallicity, rotation, and initial magnetic field strength and topology. Now that the results and conclusions of the different modeling groups agree qualitatively, we have entered a new era of precision modeling of core collapse supernovae during which the sophistication of core collapse supernova models must advance to make quantitatively accurate predictions for the outcome of massive stellar collapse---explosion or collapse---and all associated observables: explosion energies, neutron star kicks and spins, nucleosynthesis, neutrino and gravitational wave emissions, etc. For recent reviews detailing the progress made to date, we refer the reader to \citet{Mueller20}, \citet{MeEnMe20}, \citet{BuVa21}, \citet{YaNaAk24}, and \citet{Janka25}.

In the near term, the advancement of core collapse supernova modeling will require developments on several fronts, across {\em all} modeling efforts (some efforts have already implemented one or more of the following): (1) The implementation of fully general relativistic treatments of gravity, hydrodynamics/magnetohydrodynamics, and neutrino kinetics. (2) The implementation of three-dimensional neutrino kinetics. (3) The development of closures for two-moment neutrino kinetics that satisfy Fermi--Dirac statistics. (4) The development of discretizations of the equations of general relativistic neutrino radiation hydrodynamics and magnetohydrodynamics that simultaneously conserve lepton number and energy. (5) The inclusion of nuclear burning and the proper treatment of stellar matter not in nuclear statistical equilibrium (NSE). 
In the longer term, advancement will require grappling with the challenging problems, to say the least, of neutrino flavor mixing, the interaction of neutrinos with bulk nuclear matter, and the nucleon--nucleon interaction underpinning the bulk nuclear matter equation of state, obviously involving in some cases communities---e.g., nuclear physics theory---outside the core collapse supernova modeling community.

\begin{figure}
 \centering
        \includegraphics[width=0.5\textwidth]{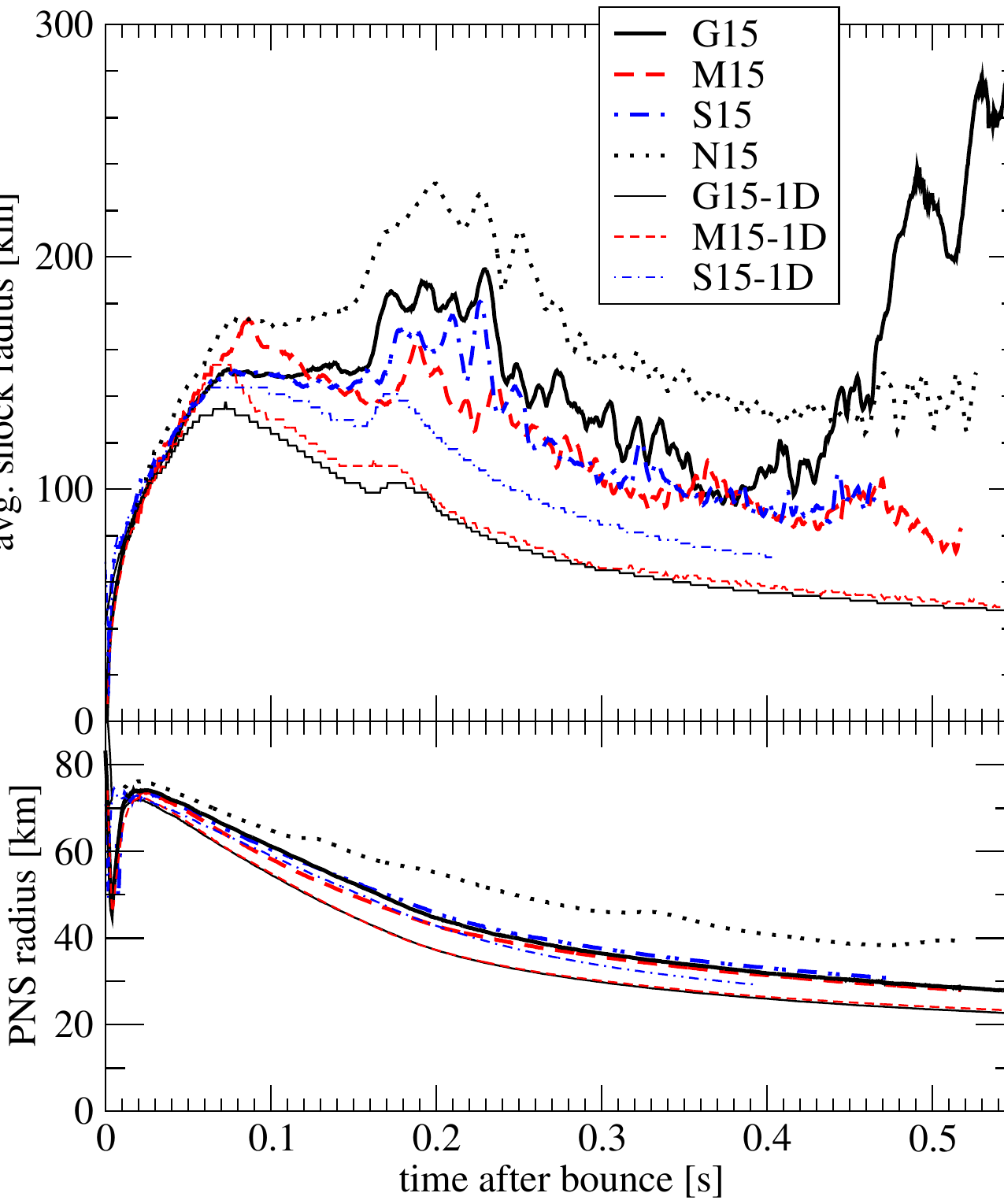}
 \caption{Head to head comparison from \citet{MuJaMa12} of one- and two-dimensional simulations using an effective potential approach (M15) versus general relativity (G15). Plotted are the angle-averaged shock trajectories.}
\label{fig:EffPotvsGR}
\end{figure}

\section{The Near Term}

\subsection{The Development of General Relativistic Core Collapse Supernova Models}

The majority of the progress made over the past decade, bringing us to an historic point in core collapse supernova modeling and theory, was made by a variety of simulation frameworks that approximated the impact of general relativistic gravity. The most widely used approach, first proposed by \citet{RaJa02} and later developed by \citet{MaDiJa06}, corrects for strong-field gravity in the context of Newtonian hydrodynamics as follows.

The Newtonian equation for hydrostatic equilibrium in spherical symmetry is given by the simple equation
\begin{equation}
\frac{1}{\rho}\frac{\partial P}{\partial r}=-\frac{\partial \Phi_{\rm Newtonian}}{\partial r},
\label{eq:NewtonianHydrostaticEquilibrium}
\end{equation}
where
\begin{equation}
\Phi_{\rm Newtonian}(r)=-4\pi \int_{0}^{\infty} dr^{'}(r^{'})^{2}\frac{\rho}{|r-r^{'}|}
\label{eq:NewtonianPotential}
\end{equation}
and $\rho$ is the rest-mass density.
On the other hand, general relativistic hydrostatic equilibrium in spherical symmetry is given by the Tolman--Oppenheimer--Volkov (TOV) equation

\begin{equation}
\frac{\partial P}{\partial r}=-\frac{G}{r^2}(\rho+\frac{P}{c^2})(m+4\pi r^{3}\frac{P}{c^2})(1-\frac{2Gm}{c^{2}r})^{-1}.
\label{eq:TOV}
\end{equation}
\citet{RaJa02} recognized that Eqn. (\ref{eq:TOV}) can be recast in the form

\begin{equation}
\frac{1}{\rho}\frac{\partial P}{\partial r}=-\frac{\partial \Phi_{\rm TOV}}{\partial r}.
\label{eq:NHEwithTOVPotential}
\end{equation}
where

\begin{equation}
\frac{\partial \Phi_{\rm TOV}}{\partial r}
\equiv
\frac{1}{\rho}\frac{G}{r^2}(\rho+\frac{P}{c^2})(m+4\pi r^{3}\frac{P}{c^2})(1-\frac{2Gm}{c^{2}r})^{-1}
\label{eq:TOVPotential}
\end{equation}
or

\begin{equation}
\Phi_{\rm TOV}(r)
=
-4\pi
\int_{r}^{\infty}dr'\frac{1}{\rho}\frac{G}{(r')^2}(\rho+\frac{P}{c^2})
(\frac{m}{4\pi}+(r')^{3}\frac{P}{c^2})
(1-\frac{2Gm}{c^{2}r'})^{-1}.
\label{eq:TOVPotentialII}
\end{equation}
The effective potential approach is implemented in the context of three-dimensional models as follows. Given the Newtonian gravitational potential in three dimensions

\begin{equation}
\Phi_{\rm Newtonian}(\vec{r})=-\int_{0}^{\infty} d^{3}r^{'}\frac{\rho}{|\vec{r}-\vec{r}^{'}|},
\label{eq:NewtonianPotential3D}
\end{equation}
an effective potential is defined as

\begin{equation}
\Phi_{\rm Effective}(\vec{r})\equiv \Phi(\vec{r})-\Phi_{\rm Newtonian}(r)+\Phi_{\rm TOV}(r),
\label{eq:EffectivePotential}
\end{equation}
where $\Phi_{\rm Newtonian}(r)$ and $\Phi_{\rm TOV}(r)$ are defined by Eqns. (\ref{eq:NewtonianPotential}) and (\ref{eq:TOVPotentialII}), respectively. To compute $\Phi_{\rm Newtonian}(r)$ and $\Phi_{\rm TOV}(r)$, the three-dimensional simulation data are first spherically averaged.

\begin{figure}
 \centering
        \includegraphics[width=0.5\textwidth]{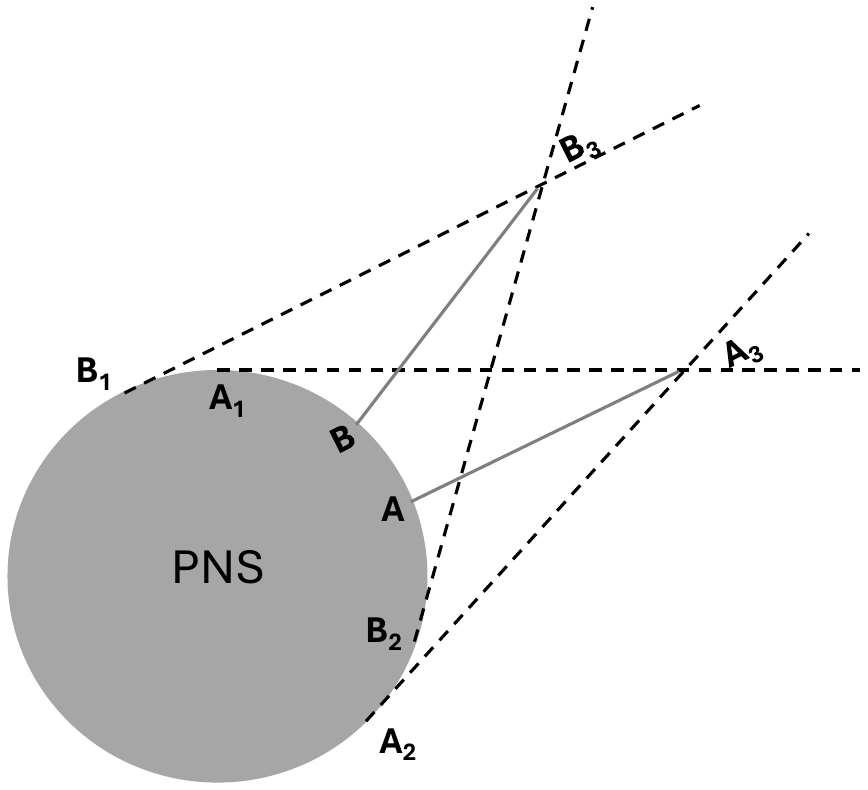}
 \caption{Illustration of ray-by-ray neutrino kinetics. Two spherically symmetric problems are shown. Neutrino heating at $\rm{A}_{3}$ is computed as if the thermodynamic conditions at $\rm{A}$ are the same over the entire surface between $\rm{A}_{1}$ and $\rm{A}_{2}$. Similarly, neutrino heating at $\rm{B}_{3}$ is computed as if the thermodynamic conditions at $\rm{B}$ are the same over the entire surface between $\rm{B}_{1}$ and $\rm{B}_{2}$. The more the thermodynamic conditions differ at points $\rm{A}$ and $\rm{B}$, the more the neutrino heating at points $\rm{A}_{3}$ and $\rm{B}_{3}$ will be overestimated or underestimated.}
\label{fig:RbRIllustration}
\end{figure}

A direct comparison between on the one hand a general relativistic treatment of gravity, hydrodynamics, and neutrino kinetics and on the other the use of the effective potential, Eqn. (\ref{eq:EffectivePotential}), Newtonian hydrodynamics, and general relativistic neutrino kinetics restricted for consistency to include only gravitational redshift and time dilation were performed by \citet{LiRaJa05}. They found that the use of an effective potential overestimates the impact of a general relativistic treatment of gravity. Motivated by this, modifications of the effective potential approach were proposed by \citet{MaDiJa06}.

The only head-to-head comparison of the results of multidimensional simulations implementing an effective potential versus general relativity was conducted by \citet{MuJaMa12}. Their two-dimensional model, M15, implemented an effective potential for gravity, Newtonian hydrodynamics, and redshift and time dilation in the ray-by-ray neutrino kinetics, consistent with the use of an effective potential. Their model, G15, implemented general relativistic gravity, hydrodynamics, and ray-by-ray neutrino kinetics, all in the xCFC approximation. Considering one measure of the differences found, at the end of the M15 run the average shock radius was approximately 80 km, whereas at the end of the G15 run the average shock radius was approximately 280 km (see Figure \ref{fig:EffPotvsGR}).

The majority of sophisticated three-dimensional models published to date implemented this effective potential approach to approximate the strong-field gravity of general relativity. While the approach is well informed, its efficacy, like with all approximations, had to be assessed by comparing with the results of models that implement general relativistic treatments of all model components. Nothing can replace the latter, and the advancement of core collapse supernova theory requires that approximations to general relativistic gravity, hydrodynamics, and neutrino kinetics be eliminated. The results of \citet{MuJaMa12} confirm this. This has been accomplished recently (using an extensive set of weak interactions) by \citet{Kuroda2021}. 

\subsection{The Implementation of Three-Dimensional Neutrino Kinetics}

As in the case of the use of an effective gravitational potential to approximate the effects of strong gravity, several groups that have contributed to the progress to date have implemented a ray-by-ray approximation when including neutrino kinetics in the models. In the ray-by-ray approach, the neutrino kinetics for each $(\theta,\phi)$ in the three-dimensional domain is evolved assuming spherical symmetry, as illustrated in Figure \ref{fig:RbRIllustration}. 

\begin{figure*}
\includegraphics[width=0.5\textwidth]{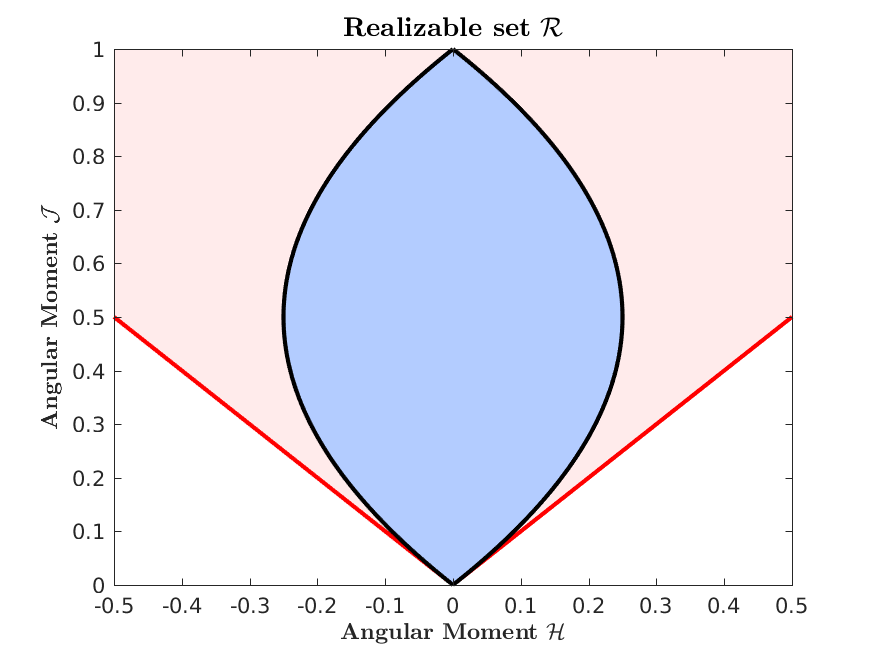}\hfill
\includegraphics[width=0.5\textwidth]{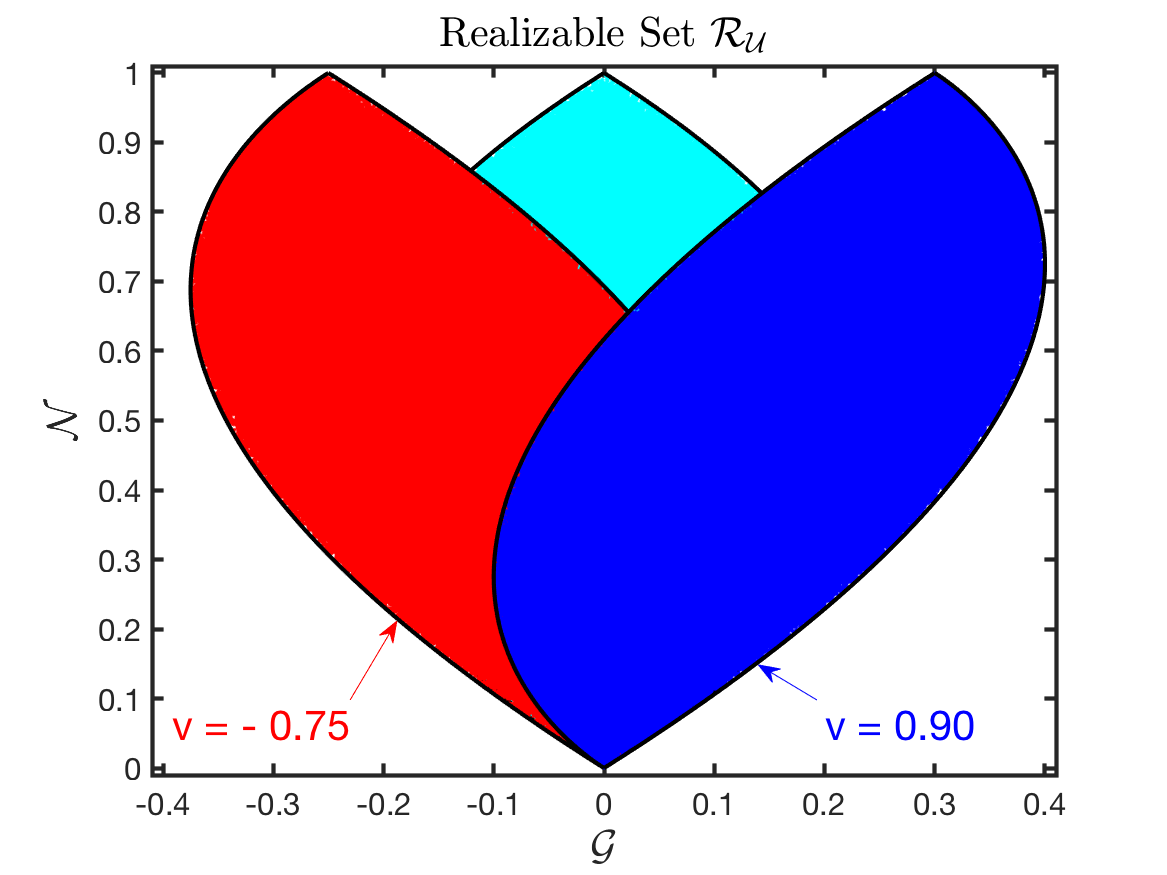}\\
\caption{Realizable domains for Maxwell--Boltzmann statistics [left panel \citep{ChEnHa19}] and Fermi--Dirac statistics (right panel). In the left panel, the realizable domain is shaded in red, where the flux, $\mathcal{H}$, is related to the energy density, $\mathcal{I}$, by $\mathcal{H}\le \mathcal{I}$. Also included in the left panel is the realizable domain for zero fluid velocity and Fermi--Dirac statistics. The zeroth angular moment is bounded by the fact that  the distribution function is bounded: $f\in [0,1]$. When the distribution function is $1$ for all angles, the zeroth moment is maximum whereas the first moment is zero (the flux is zero when the distribution is isotropic in angle). The right panel illustrates three realizable domains for three different fluid velocities: $v=-0.75$, $v=0$, and $v=0.90$, clearly illustrating the velocity dependence of realizability.}
\label{fig:RealizableDomains}
\end{figure*}

This approach was first proposed by \citet{RaJa02} and was motivated in part by practical considerations. Sophisticated neutrino kinetics solvers for spherically symmetric simulations were available given the previous development of spherically symmetric models of core collapse supernovae. The ray-by-ray approach is exact, yet redundant, in spherical symmetry. The efficacy of the approximation thus stems from the degree to which the neutrino source---i..e., the proto-neutron star---is spherically symmetric {\em in a temporally-averaged sense}. A comparison of the results of three-dimensional core collapse supernova simulations using both ray-by-ray and three-dimensional neutrino kinetics was conducted by \citet{GlJuJa19}. They indeed found that the time-averaged results from both neutrino kinetics treatments were in good agreement. This of course is an important validation of the outcomes of the three-dimensional models performed to date using the ray-ray-ray treatment, which were instrumental in advancing the field to its present state. Nonetheless, as before, a three-dimensional treatment was needed to assess the efficacy of the ray-by-ray approach and was done so, albeit successfully, in a limited comparison. Moreover, sixty years of modeling has demonstrated that the outcomes of simulations are sensitive to (i) the physics included (the underlying equations used for the macroscopic evolution of the explosion dynamics and the input microphysics used, which comprises the inclusion of a complete set of weak interactions demonstrated to be important to simulation outcomes, the most advanced treatments of these interactions, and the use of a nuclear equation of state that remains admissible based on nuclear experiment and astronomical observations of neutron star masses and radii), (ii) the numerical methods adopted for the solution of the equations adopted, especially any approximations made at this level in order to reduce the computational cost, and (iii) numerical resolution. It would be impossible to know in all cases going forward whether or not the ray-by-ray approach is sufficient to the level of accuracy required, which may vary case by case, particularly for borderline cases. Thus, the assumption that a ray-by-ray approach will be sufficient for all future modeling efforts has limited support.

In addition to the simulations by \citet{GlJuJa19} described above, three-dimensional neutrino kinetics with an extensive set of neutrino weak interactions has been implemented by \citet{VaBuRa18a,VaBuRa19,BuRaVa19} and \citet{Kuroda2021}, in the former cases using the effective potential approach and in the latter case in general relativity. 

\begin{figure}
  \centering
  \begin{tabular}{cc}
    \includegraphics[width=0.5\textwidth]{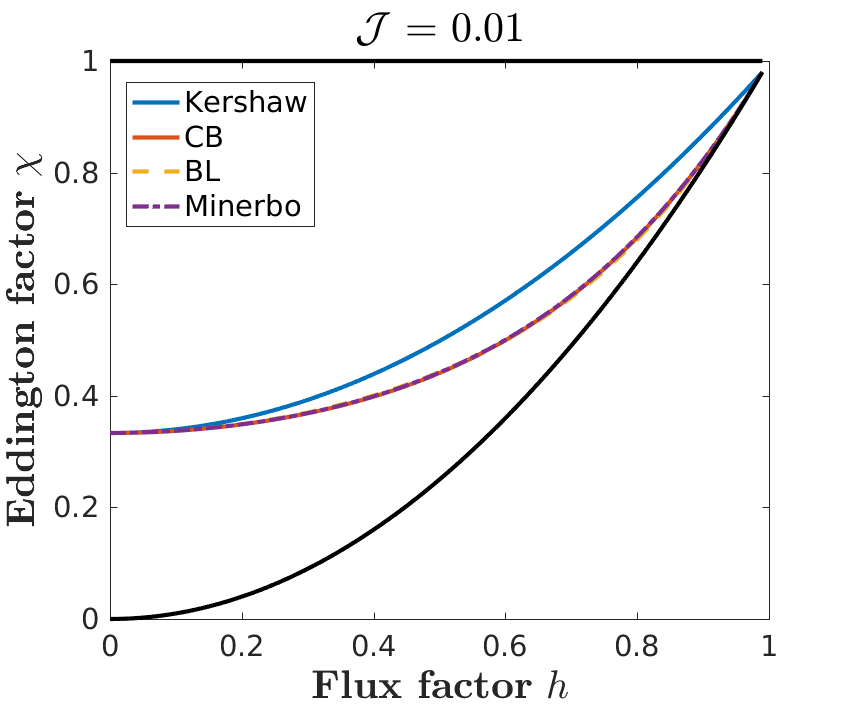}
    \includegraphics[width=0.5\textwidth]{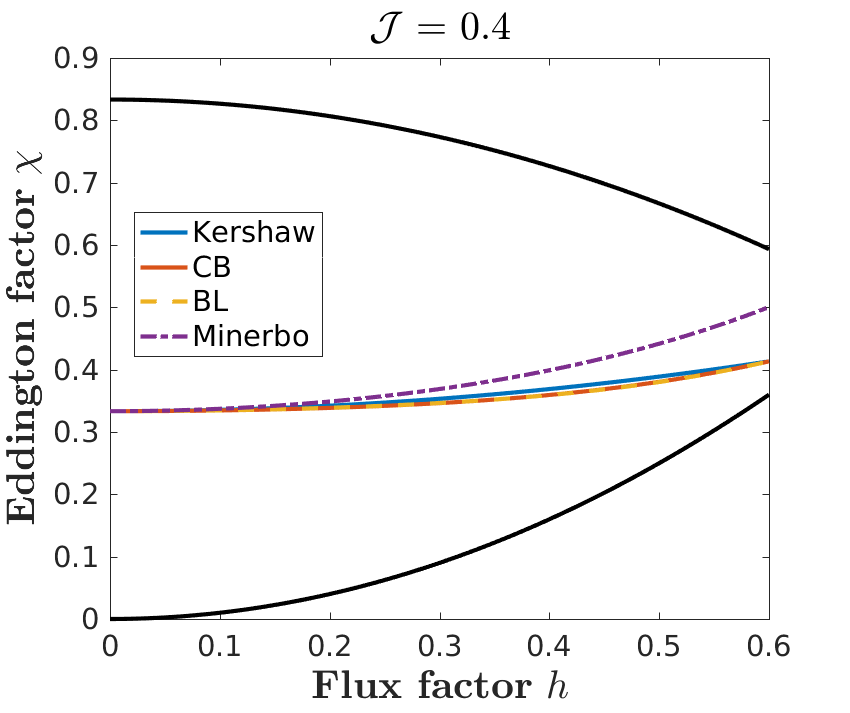} \\
    \includegraphics[width=0.5\textwidth]{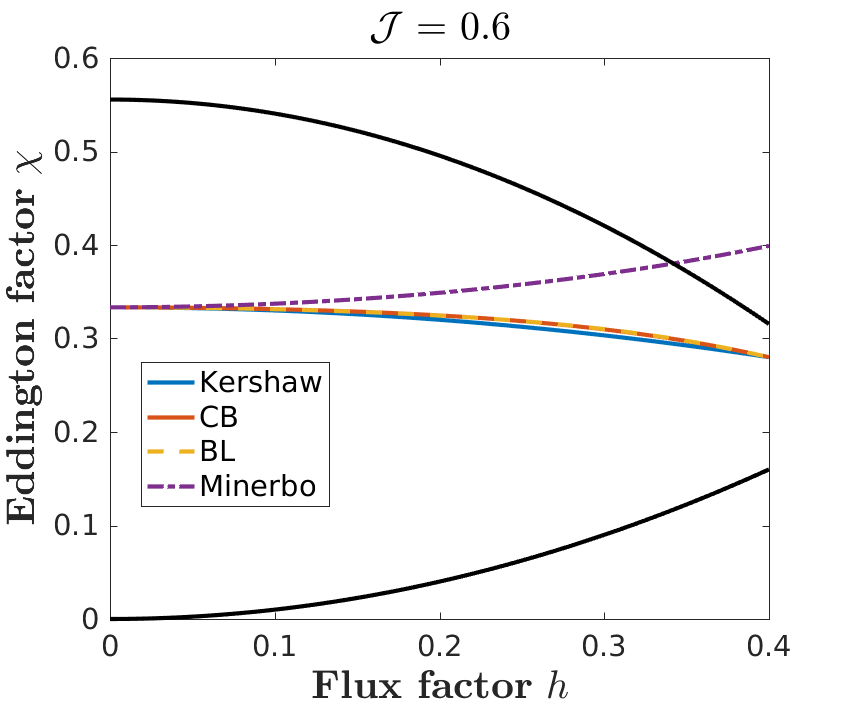}
    \includegraphics[width=0.5\textwidth]{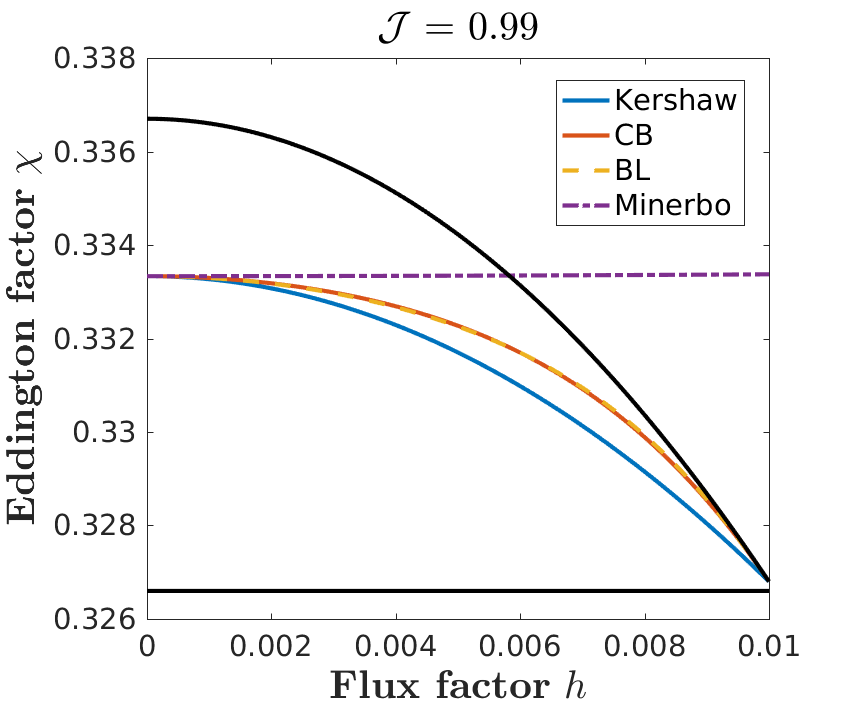}
  \end{tabular}
   \caption{Eddington factor versus flux factor for four different values of the number density \citep{ChEnHa19} In each case, four different closures are considered. The upper and lower bounds, Eqn. (\ref{eq:eddingtonFactorBounds}), are plotted (solid black curves), as well. At low number density (upper left panel), Minerbo closure agrees with closures that derive from Fermi--Dirac statistics [e.g., Chernohorsky--Bludman (CB) closure] and remains within the bounds. As the number density increases, the results from Minerbo closure deviate from the results from CB closure and the Eddington factor no longer respects the bounds for all values of the flux factor.}
  \label{fig:EddingtonFactorswithDifferentClosure}
\end{figure}

\subsection{Two-Moment Neutrino Kinetics: The Closure Problem}

Efforts by \citet{IwOkNa20} to solve the general relativistic Boltzmann equation for each neutrino species in the context of three-dimensional models have begun, albeit with limited applicability given the computational cost of such an approach to neutrino kinetics in core collapse supernovae, which remains well beyond even the most capable of present-day leadership-class supercomputers. In light of the prohibitive computational cost of Boltzmann kinetics, three-dimensional general relativistic simulations have been conducted using two-moment neutrino kinetics by \citet{Kuroda2021}. In the two-moment approach adopted in this work, the evolution equations for the spectral zeroth and first angular moments of the neutrino radiation field, as measured by an Eulerian observer, $E_{\varepsilon}$ and $F^{\alpha}_{\varepsilon}$, respectively, are

\begin{eqnarray}
\partial_t \sqrt{\gamma}E_{\varepsilon}+\partial_i \sqrt{\gamma}(\alpha F_{\varepsilon}^i-\beta^i E_{\varepsilon})
+\sqrt{\gamma}\alpha \partial_\varepsilon \bigl(\varepsilon 
M^{\mu\alpha\beta}_{\varepsilon}(\nabla_\beta u_\alpha) n_\mu\bigr)  =
\sqrt{\gamma}(\alpha P^{ij}_{\varepsilon}K_{ij}-F_{\varepsilon}^i\partial_i \alpha-\alpha S_{\varepsilon}^\mu n_\mu)
\label{eq:ZerothMoment}
\end{eqnarray}
and

\begin{eqnarray}
&&
\partial_t \sqrt{\gamma}{F_{\varepsilon}}_i+\partial_j \sqrt{\gamma}(\alpha 
{P_{\varepsilon}}_i^j-\beta^j {F_{\varepsilon}}_i)
-\sqrt{\gamma}\alpha \partial_\varepsilon\bigl(\varepsilon 
\gamma_{i\mu}M^{\mu\alpha\beta}_{\varepsilon}\nabla_\beta u_\alpha
\bigr)=
\nonumber \\
&&
\sqrt{\gamma}[-E_{\varepsilon}\partial_i\alpha +{F_{\varepsilon}}_j\partial_i \beta^j+(\alpha/2) 
P_{\varepsilon}^{jk}\partial_i \gamma_{jk}+\alpha S^\mu_{\varepsilon} \gamma_{i\mu}].
\label{eq:FirstMoment}
\end{eqnarray}
Here $\varepsilon$ is the neutrino energy measured in the comoving frame. The quantities $\alpha$, $\beta^i$, $\gamma^{ij}$, and $K_{ij}$ are the usual $3+1$ variables---the lapse function, shift vector, three-metric, and extrinsic curvature, respectively. $\gamma\equiv{\rm det}(\gamma_{ij})$ is the determinant of the three-metric. Most important for our discussion here: $P^{ij}$ and $M^{\mu\alpha\beta}_{\varepsilon}$ are the second and third angular moments of the neutrino radiation field, respectively, both measured in the Eulerian frame, as in the cases of the zeroth and first moments.

\begin{figure}
 \centering
        \includegraphics[width=0.65\textwidth]{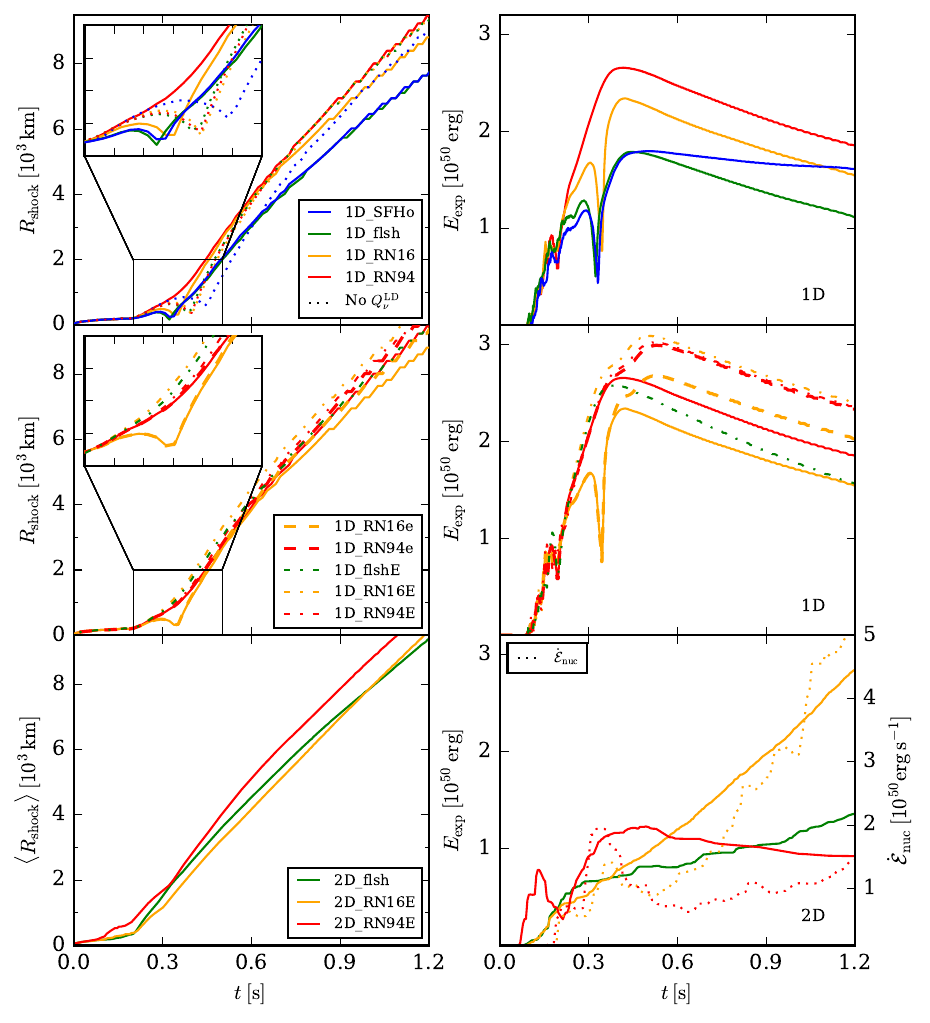}
 \caption{Plots of average shock radii (left panels) and explosion energies (right panels) for the one- and two-dimensional models of \citet{NaReOb23} using different treatments of nuclear burning in non-NSE matter---in particular, nuclear networks with 16 and 94 nuclear species. Of particular note are the plots corresponding to their two-dimensional models.}
\label{fig:AvgShockTraj}
\end{figure}

As with the hydrodynamics equations, which must be closed using an equation of state relating the pressure to the variables evolved, Eqns. (\ref{eq:ZerothMoment}) and (\ref{eq:FirstMoment}) require closure, as well. The second and third neutrino radiation field moments must be related to the first two. This is the ``closure problem." Closure is typically implemented in the comoving frame, where the relationship between the second and zeroth moments of the neutrino radiation field are well known in the diffusion and free-streaming limits---e.g., in the former case the ratio of the components of the second moment to the zeroth moment is $1/3$.

Closures can be calculated or prescribed. In the former case, the Boltzmann equation or some approximation thereof is solved, and the resulting distribution function can be used to calculate all necessary moments not evolved in Eqns. (\ref{eq:ZerothMoment}) and (\ref{eq:FirstMoment}). In the latter case, an analytic expression is assumed. 

Particle statistics enter in the prescription of an analytic closure. To date, the prescriptions adopted assume Maxwell--Boltzmann statistics, not Fermi--Dirac statistics. In the case of Maxwell--Boltzmann statistics, the bounds on the evolved neutrino radiation field moments are simple and independent of the velocity of the fluid. The first moment is bound by the zeroth moment (in units where $c=1$). For Fermi--Dirac statistics, the bounds on these two evolved moments are much more complex and depend on the fluid velocity. An example of the difference in the ``realizable'' domains, where the moments satisfy the bounds, is shown in Figure \ref{fig:RealizableDomains}.

\citet{ChEnHa19} investigated the appropriateness of closing a two-moment system using a closure based on Maxwell--Boltzmann statistics. They began with the nonrelativistic Boltzmann equation 

\begin{equation}
  \partial_{t} f +\vec{\ell}\cdot\nabla f
  =\frac{1}{\tau}\,\mathcal{C}(f),
  \label{eq:boltzmann}
\end{equation}
where the distribution function $f(\omega,\varepsilon,\vec{x},t)$ gives the number of neutrinos propagating in the direction $\omega=(\theta,\phi)~|~\theta\in[0,\pi],\phi\in[0,2\pi)$, with energy $\varepsilon$, at position $\vec{x}$ and time $t$. Spherical momentum-space coordinates $(\varepsilon,\omega)$ are used, and the unit vector $\vec{\ell}(\omega)$ is parallel to the neutrino three-momentum $\vec{p}=\varepsilon\,\vec{\ell}$. On the right-hand side of Equation (\ref{eq:boltzmann}), $\tau$ is the ratio of the neutrino mean-free path to some characteristic length scale of the problem. In opaque regions, $\tau\ll1$, while for free streaming particles, $\tau\gg1$. The spectral angular moments of the distribution function are given by

\begin{equation}
  \big\{\,\mathcal{J},\vec{\mathcal{H}},\vec{\mathcal{K}}\,\big\}(\vec{x},\varepsilon, t)
  =\frac{1}{4\pi}\int f(\vec{x}, t, \omega,\varepsilon)\,\{\,1,\vec{\ell},\vec{\ell}\otimes\vec{\ell}\,\}\,d\omega.  
  \label{eq:angularMoments}
\end{equation}
We refer to $\mathcal{J}$ (zeroth moment) as the neutrino density, $\vec{\mathcal{H}}$ (first moment) as the neutrino flux, and $\vec{\mathcal{K}}$ (second moment) as the neutrino stress tensor.

\begin{figure}
 \centering
        \includegraphics[width=0.65\textwidth]{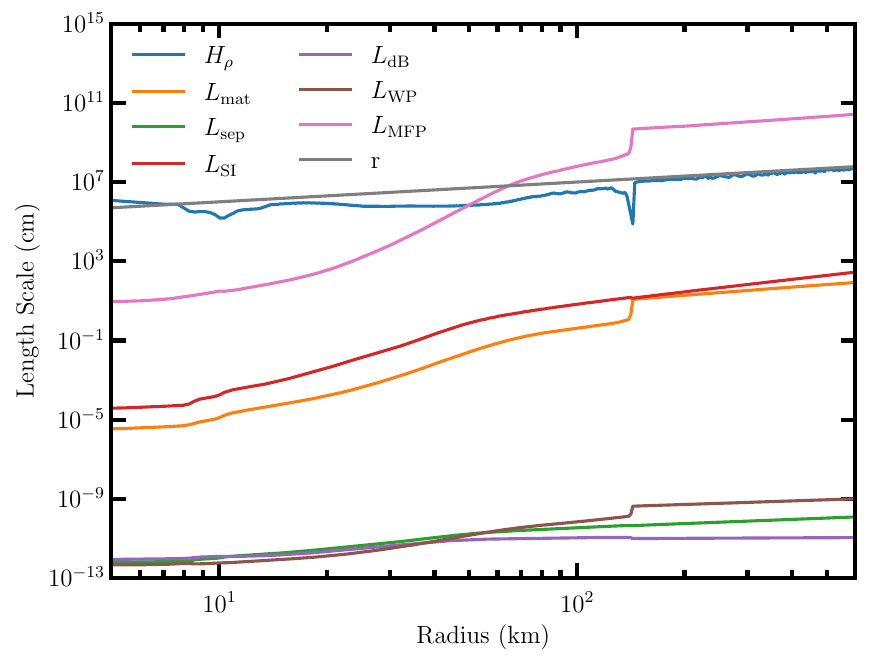}
 \caption{The length scale, $L_{\rm SI}$, is the length scale corresponding to neutrino--neutrino forward scattering, responsible for fast flavor transformation in core collapse supernovae, plotted here as a function of radius at the time of maximum shock radius in the one-dimensional simulation from which the data were drawn \citep{JoRiWu25}. On the other hand, the length scale, $H_{\rho}$, is the density scale height, a length scale that is resolved in numerical simulations. In the region below $10^{2}$ km, which contains the proto-neutron star in this case at this time, $L_{\rm SI}$ and $H_{\rho}$ differ by at least four orders of magnitude, the difference approaching at least ten orders of magnitude as $r\rightarrow 0$.}
\label{fig:lengthscales}
\end{figure}

Taking the zeroth and first angular moments of Eqn. (\ref{eq:boltzmann}) yields the two-moment system

\begin{equation}
  \partial_{t}\vec{\mathcal{M}}+\nabla\cdot\vec{\mathcal{F}}=\frac{1}{\tau}\,\vec{\mathcal{C}}(\vec{\mathcal{M}}),
  \label{eq:momentEquations}
\end{equation}
where $\vec{\mathcal{M}}=(\mathcal{J},\vec{\mathcal{H}})^{T}$ and $\vec{\mathcal{F}}=(\vec{\mathcal{H}},\vec{\mathcal{K}})^{T}$. The system is closed by relating the second angular moment, $\vec{\mathcal{K}}$, to the first two angular moments, $\mathcal{J}$ and $\vec{\mathcal{H}}$. Defining the Eddington tensor

\begin{equation}
  \vec{k}\equiv\frac{\vec{\mathcal{K}}}{\mathcal{J}}
  \label{eq:EddingtonTensor}
\end{equation}
and assuming the radiation field is axisymmetric about  

\begin{equation}
\hat{h}=\vec{\mathcal{H}}/|\vec{\mathcal{H}}|,
\label{eq:preferreddirection}
\end{equation}
following \citet{Levermore84} \citet{ChEnHa19} write

\begin{equation}
  \vec{k}=\frac{1}{2}\big[\,\big(1-\chi\big)\,\vec{I}+\big(3\,\chi-1\big)\,\hat{h}\otimes\hat{h}\,\big],
  \label{eq:eddingtonTensor}
\end{equation}
where $\chi=\chi(\mathcal{J},|\vec{\mathcal{H}}|)$ is the Eddington factor.  

\citet{Levermore84} and \citet{LaBa11} demonstrated that realizability of the moment triplet $(\mathcal{J},\vec{\mathcal{H}},\vec{\mathcal{K}})$ (with $\vec{\mathcal{K}}$ given by Eqn. (\ref{eq:eddingtonTensor})) is equivalent to the following requirement for the Eddington factor
\begin{equation}
  \chi_{\mbox{\tiny min}}
  =\max\big(1-\frac{2}{3\mathcal{J}},h^{2}\big)
  <\chi<\min\big(1,\frac{1}{3\mathcal{J}}-\frac{\mathcal{J}}{1-\mathcal{J}}h^{2}\big)=\chi_{\mbox{\tiny max}}.  
  \label{eq:eddingtonFactorBounds}
\end{equation}
In Figure \ref{fig:EddingtonFactorswithDifferentClosure}, the Eddington factor, $\chi$, is plotted against the flux factor, $h$, for four different closures and four different values of the number density, $\mathcal{J}$. Bounds on the Eddington factor, $\chi_{\tiny\rm min}$ and $\chi_{\tiny\rm max}$, are provided (black curves). The Minerbo closure, the most widely used in three-dimensional simulations that have implemented two-moment neutrino kinetics, assumes Maxwell--Boltzmann statistics. At low occupation numbers, all four closures remain within the bounds and there is good agreement between the results using Minerbo closure and Chernohorsky and Bludman (CB) closure for Fermi--Dirac statistics. However, as the occupation numbers increase, the Minerbo closure violates the bounds, already at intermediate occupancy. This suggests that these bounds may be violated across much of the proto-neutron star in a core collapse supernova simulation. Thus, closures that satisfy Fermi--Dirac statistics must be found and implemented.

\begin{figure}
\begin{center}
\includegraphics[scale=0.5]{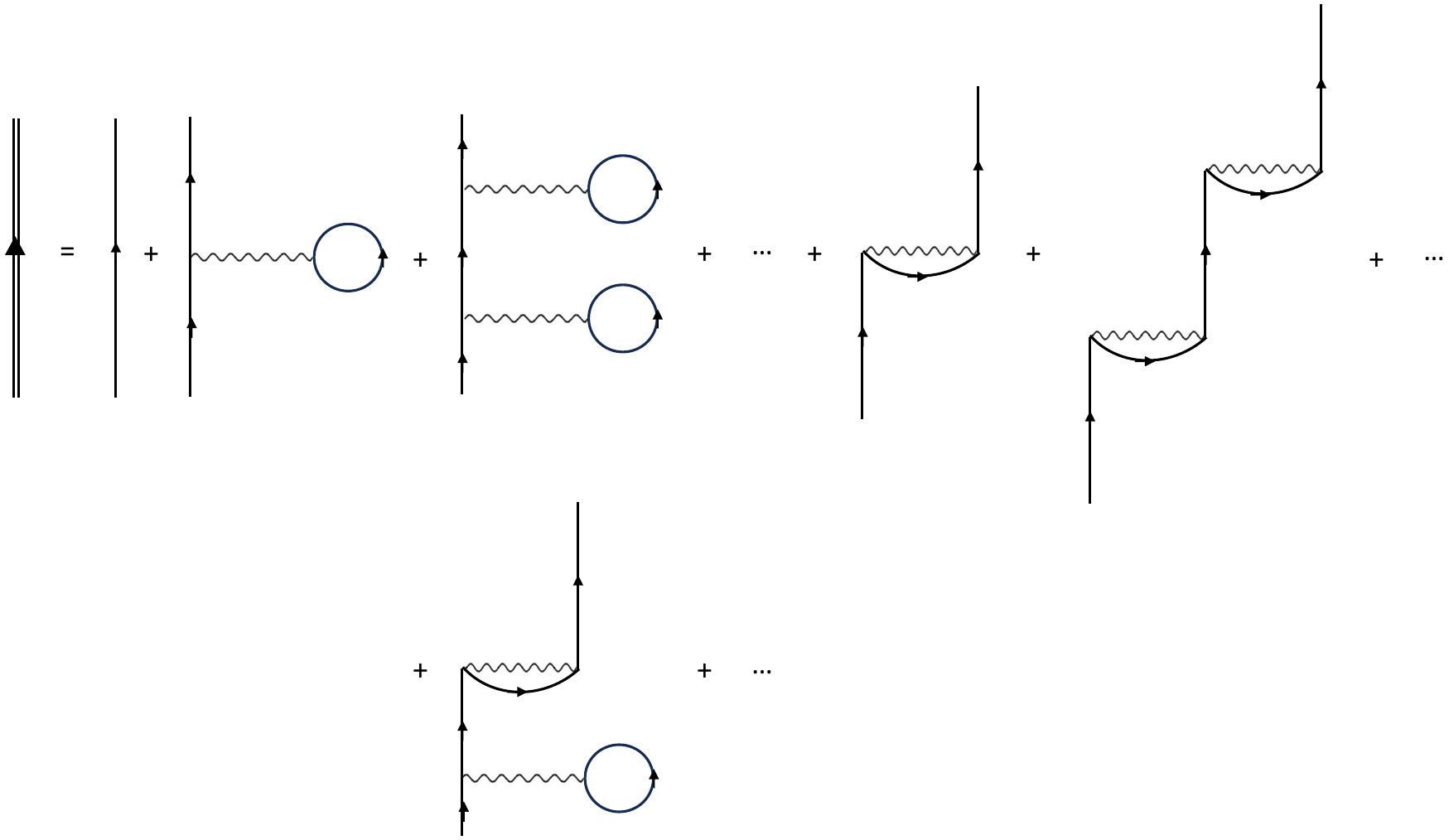}
\end{center}
\vspace{-15pt}
\caption{Diagrammatic representation of the Hartree--Fock corrections to the single-particle dispersion relations (single-particle masses) resulting from forward scattering (bubble diagrams) and exchange scattering (oyster diagrams) of a nucleon in the mean field created by all of the other nucleons in the medium.}
\label{fig:HF}       
\end{figure}

\subsection{Simultaneous Conservation of Lepton Number and Energy}

In the continuum limit, different forms of the integro-partial differential equations governing the evolution of the neutrino radiation field are equivalent. But this equivalence breaks down in the discrete limit. Discretizations of these equations must preserve physical conservation laws. In the case of core collapse supernovae, there are two: conservation of total lepton number and conservation of energy. Achieving both in the discrete limit is a significant challenge. Addressing the challenge begins in the continuum limit where formulations of the underlying integro-partial differential equations exist that are manifestly conservative for neutrino lepton number or for neutrino energy. Beginning with, for example, the number conservative formulation and discretizing it, the challenge becomes adapting the discretization to also conserve energy. With regard to this last point, there are discretizations of the underlying equations that are better suited than others. For example, when using finite differencing to discretize the continuum equations, the finite differencing of individual terms in the equations is not independent if conservation of energy is to be achieved beginning with a number conservative approach. Global conservation of energy is defined in the Eulerian frame of reference at rest with respect to the fixed stars. The equation for this globally conserved energy results from a cancellation of terms in the continuum equations (e.g., see \cite{MeEnMe20}). This cancellation must occur in the discrete case as well. Canceling terms cannot be discretized independently. Such lepton number and energy-conservative discretizations were achieved in the context of spherically symmetric, general relativistic simulations of core collapse supernovae by \citet{LiMeMe04} and in the context of multidimensional simulations using ray-by-ray neutrino kinetics by \citet{MuJaDi10}, in both cases using this discretization-matching procedure. As the number of terms proliferates in the case of three-dimensional general relativistic neutrino kinetics, the cancellation of all dependent groupings of terms becomes increasingly difficult to achieve. 
More recently, \citet{LaEnHa25} presented a finite-element method for three-dimensional, two-moment, spectral neutrino kinetics in the $\mathcal{O}(v/c)$ limit. Specifically, the method implements discontinuous Galerkin (DG) discretizations for space and neutrino energy. The simultaneous conservation of lepton number and energy is naturally achieved (in this case, to $\mathcal{O}(v/c)$) in the weak formulation of DG without the need for the discretization matching that is required in the cases of finite-difference 
methods. Thus, DG methods provide a powerful approach to the challenge of satisfying both conservation laws in the discrete limit.

\begin{figure}
\begin{center}
\includegraphics[scale=0.4]{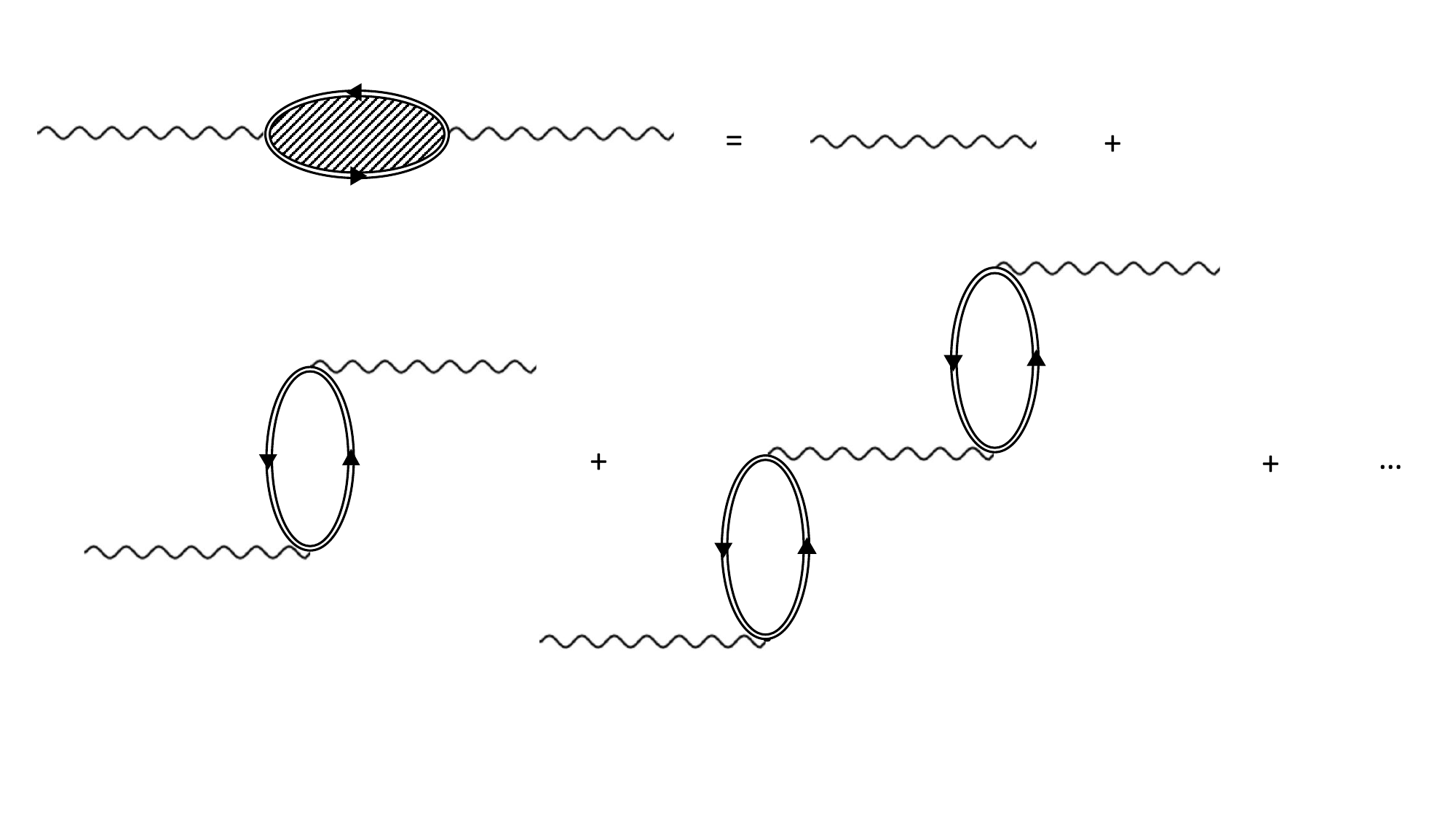}
\end{center}
\vspace{-30pt}
\caption{Sum of all ring diagrams to all orders of perturbation theory in the random phase approximation, the leading order approximation to the response of nuclear matter to a neutrino interacting with it. The interaction lines in this case are $W/Z$-mediated weak interactions, which can excite particle--hole pairs in the medium, which in turn can produce $W/Z$ bosons when deexcited, and so on, creating long-range interactions and correlations. The effective interaction is the sum of all possible sequences of particle--hole pair creation and annihilation.}
\label{fig:ringdiagramsummation}       
\end{figure}

\subsection{The Inclusion of Nuclear Kinetics}

It has now been established by \citet{NaReOb23} that proper treatment of matter not in nuclear statistical equilibrium (NSE) is important not only for explosive nucleosynthesis but for the shock dynamics as well. The inclusion of an ensemble of nuclei in non-NSE versus a single heavy nucleus has several impacts on the shock dynamics: (1) The composition in the vicinity of the shock---specifically, the number of nucleons---is altered, thereby altering the neutrino heating there. In turn, the neutrino heating lowers the ram pressure ahead of the shock, against which the shock must do work to exit the star. This facilitates explosion. (2) Nuclear burning ahead of the shock further contributes to a reduction in the ram pressure. (3) Nuclear burning behind the shock leads to an increase in the explosion energy up to tens of percent. Figure \ref{fig:AvgShockTraj} shows the comparison between three two-dimensional models conducted using the single-heavy-nucleus (SHN) approximation and two nuclear networks of 16 and 94 species. The average shock radii of the model using the SHN approximation and the 16-species network agree quite well throughout the simulations, whereas the average shock radius in the case of the 94-species network is significantly enhanced. On the other hand, the explosion energy in the case of the 16-species network is significantly enhanced relative to both the SHN approximation and the 94-species network. Thus, nuclear burning impacts the models, but the degree to which it impacts the models and the way in which it impacts the models depends on the specific nuclear network deployed, adding an additional dimension to the problem, which must be investigated.

\section{The Far Term}

\subsection{Neutrino Flavor Transformation}

The discovery of the fast flavor instability by \citet{Sawyer05} for neutrino flavor transformation and the expectation that it may occur below the region of neutrino shock-reheating in core collapse supernovae and impact shock revival has added a new level of complexity to core collapse supernova theory. Such a quantum mechanical phenomenon would require neutrino quantum kinetics able to describe both the classical phenomena of phase-space advection and neutrino--matter and neutrino--neutrino interactions and the quantum mechanical phenomenon of neutrino flavor transformation. In such a description, the neutrinos are described by the density matrix

\begin{eqnarray}
f_{ab}
=
\left(
\begin{array}{ccc}
f_{\nu_{\rm e}\nu_{\rm e}} & f_{\nu_{\rm e}\nu_{\mu}} &  f_{\nu_{\rm e}\nu_{\tau}}\\
f_{\nu_{\mu}\nu_{\rm e}} & f_{\nu_{\mu}\nu_{\mu}} &  f_{\nu_{\mu}\nu_{\tau}}\\
f_{\nu_{\tau}\nu_{\rm e}} & f_{\nu_{\tau}\nu_{\mu}} &  f_{\nu_{\tau}\nu_{\tau}}\\
\end{array}
\right)
\label{eq:densitymatrix}
\end{eqnarray}
whose diagonal elements correspond to the classical distribution functions, which give the phase-space density for each neutrino flavor---i.e., $f_{\nu_{\rm e}\nu_{\rm e}}=f_{\nu_{\rm e}}$, etc.---and whose off-diagonal elements correspond to transition ``probabilities'' between flavors. The density matrix obeys the quantum-kinetics equation \citep{RiSe22}

\begin{equation}
p^\alpha \frac{\partial f_{ab}}{\partial x^\alpha} + \frac{d p^\alpha}{d\lambda}\frac{\partial f_{ab}}{\partial p^\alpha}
= \epsilon \left({\mathcal C_{ab}}- i \left[\mathcal{H}_{ab}, f_{ab}\right]\right)
\,\ .
\label{eq:eom_ch2}
\end{equation}
where
$\epsilon = p^{\alpha}u_{\alpha}$ is the neutrino energy measured by an observing comoving with the matter, whose $4$-velocity is $u^{\alpha}$, and the Hamiltonian matrix, $\mathcal{H}$, is 

\begin{equation}
\mathcal{H}_{ab} = \mathcal{H}_\mathrm{vac} + \mathcal{H}_\mathrm{matter} + \mathcal{H}_{\nu\nu} \,\, ,
\label{eq:ham_ch2}
\end{equation}
with contributions for neutrino-mass--induced flavor transformation due to the different neutrino masses (vacuum flavor transformation), matter-induced flavor transformation due to neutrino--matter forward scattering, and neutrino-induced flavor transformation due to neutrino--neutrino forward scattering.
A flavor off-diagonal element of the density matrix for neutrinos moving in direction $\mathbf{v}$ can be decomposed into plane wave solutions 

\begin{equation}
    f^{ab}(\mathbf{x},\mathbf{p}) = \frac{f^{aa}(\mathbf{p})-f^{bb}(\mathbf{p})}{2} Q^{ab}(\mathbf{v}) e^{-iK^\alpha x_\alpha}\,\,,
\end{equation}
with real amplitude $Q^{ab}(\mathbf{v})$ and complex $4$-wave number $\mathbf{K}=(\omega,k)$. Under the assumption that $Q^{ab}(\mathbf{v})<<1$, Eqn. (\ref{eq:eom_ch2}) can be linearized, resulting in an eigenvalue equation. Complex solutions, if they exist, correspond to unstable normal modes whose amplitudes grow exponentially \citep{BaDiRa11,Volpe24}.
The linear stability analysis predicts that fast pairwise conversion of neutrinos should arise, where the number densities of electron neutrinos and antineutrinos are the same along $\vec{v}$ but not along any other direction \citep{TaSh21}.

\begin{figure}
\begin{center}
\includegraphics[scale=0.4]{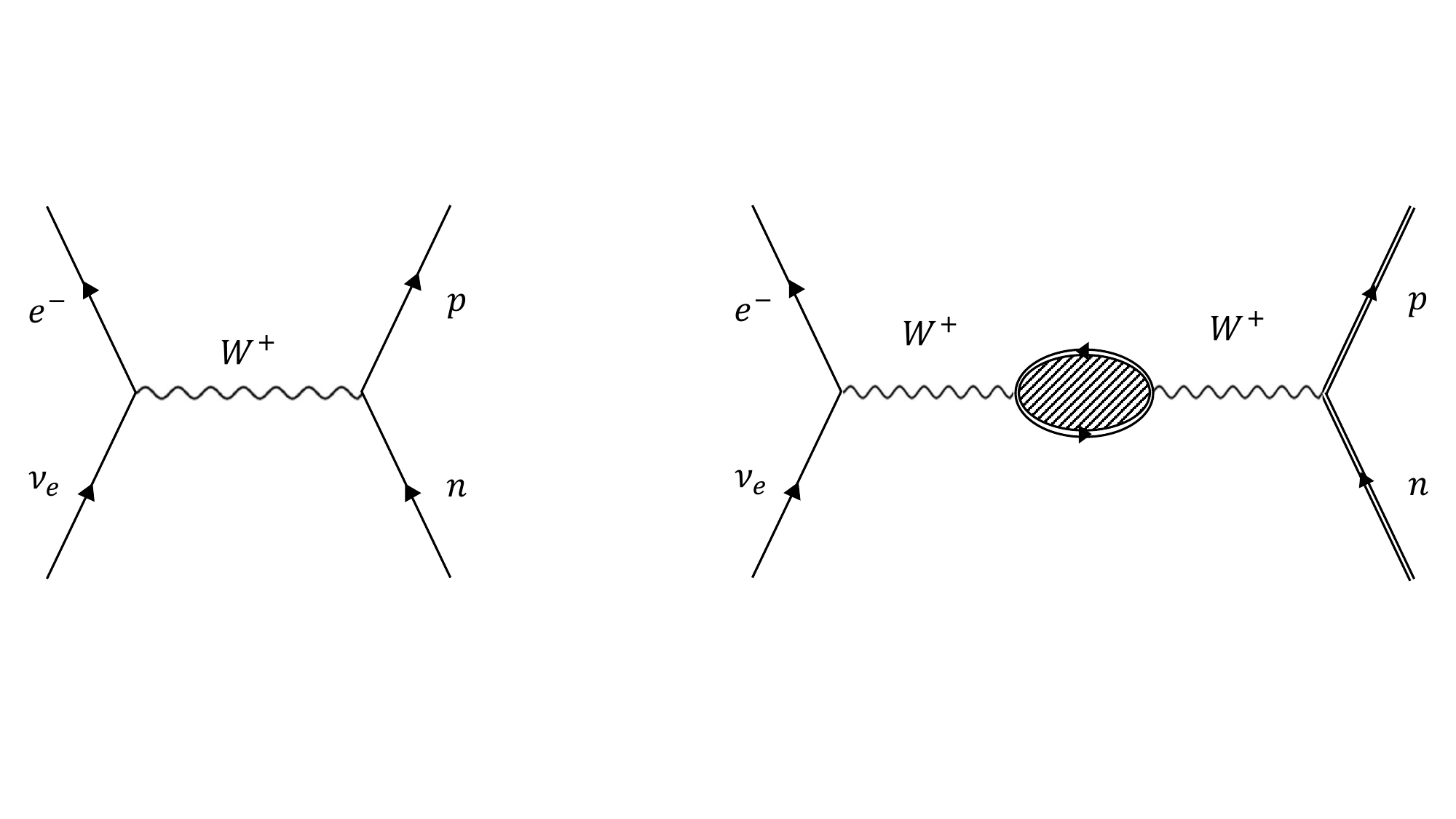}
\end{center}
\vspace{-50pt}
\caption{Electron-neutrino absorption on a free neutron via a $W^{+}$-mediated weak interaction (left panel) and on a mean-field--dressed neutron via an effective $W^{+}$-mediated weak interaction in the RPA approximation (right panel).}
\label{fig:electronneutrinoabsorption}       
\end{figure}

With the potential for fast flavor transformation and its potential implications for neutrino shock reheating now demonstrated and somewhat understood, the community turned its attention to implementing it in core collapse supernova simulations. It should be clear that a solution of Eqn. (\ref{eq:eom_ch2}) is not yet possible. As discussed earlier, even Eqn. (\ref{eq:boltzmann}) cannot be used as the sole description of neutrino kinetics in core collapse supernova simulations due to the associated computational cost. Eqn. (\ref{eq:eom_ch2}) presents a much greater challenge. Aside from the obvious addition of off-diagonal distribution functions and additional terms in the quantum kinetics equation relative to the Boltzmann kinetics equation, as demonstrated in Figure \ref{fig:lengthscales} the severest challenge arises from the fact that the spatial and temporal scales over which fast flavor transformation occurs are orders of magnitude smaller than what can possibly be resolved in numerical simulations \citep{JoRiWu25}. This necessitates, as in other astrophysical applications with directly unresolvable scales (e.g., the resolution of the turbulent flame front in Type Ia supernova simulations), the development of  subgrid models. To add insult to injury, there is an ongoing debate as to whether or not a mean field approach, with the neutrino distribution functions making up the density matrix serving as the mean fields in this case, is even appropriate \citep{LaRi25}. Some argue that an $N$-body approach is appropriate. Of course, with $\mathcal{O}(10^{58})$ neutrinos produced over $\mathcal{O}(10)$ seconds during the supernova and subsequent cooling and formation of a neutron star, it is hard to see how such an approach could be viable. Comparisons between mean-field and $N$-body approaches (albeit at small $N$) to neutrino flavor transformation in these environments have been provided by, for example, \citet{LaRi25}. They demonstrated that the fast flavor instability (in the more realistic spatially inhomogeneous case) can be disrupted by many-body correlations and that many-body effects can occur before mean-field instabilities are able to saturate. These results have profound implications regarding the feasibility of capturing fast flavor transformation physics in simulations.

\begin{figure*}[htp]
	\centering
	\includegraphics[width=0.45\textwidth]{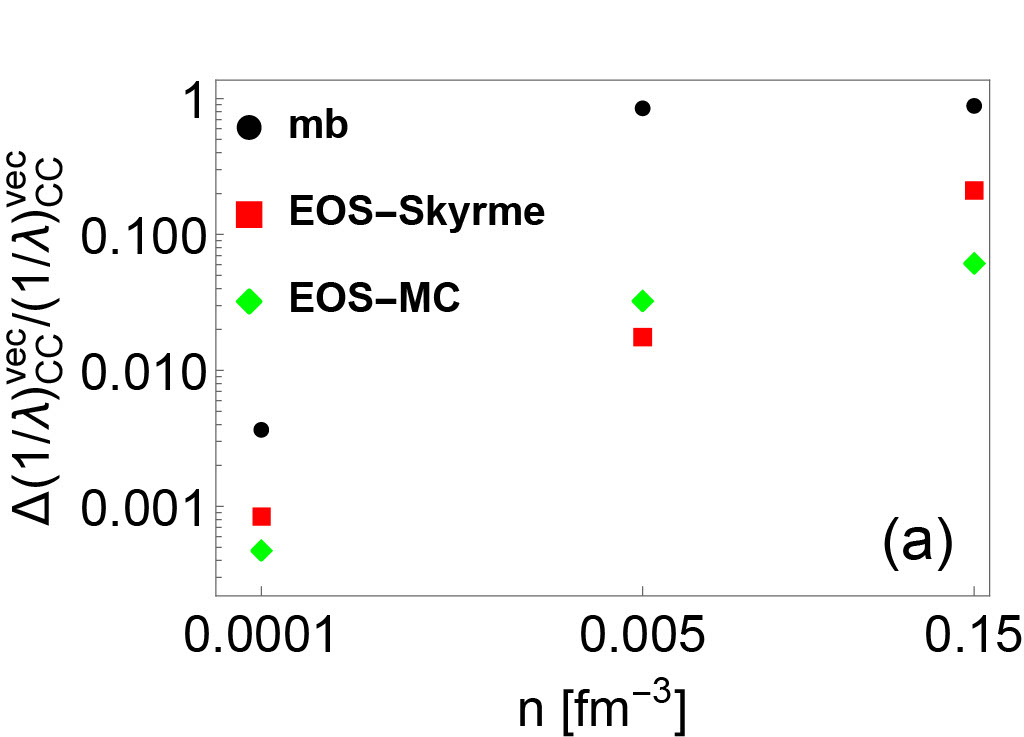}
	\includegraphics[width=0.45\textwidth]{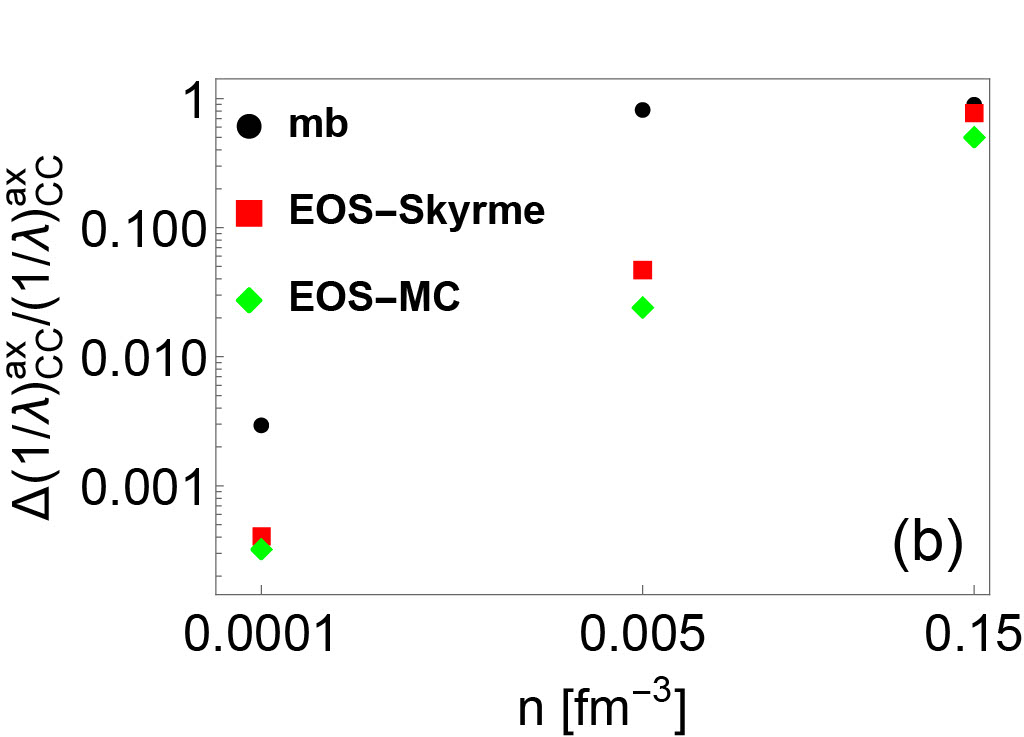}
	\includegraphics[width=0.45\textwidth]{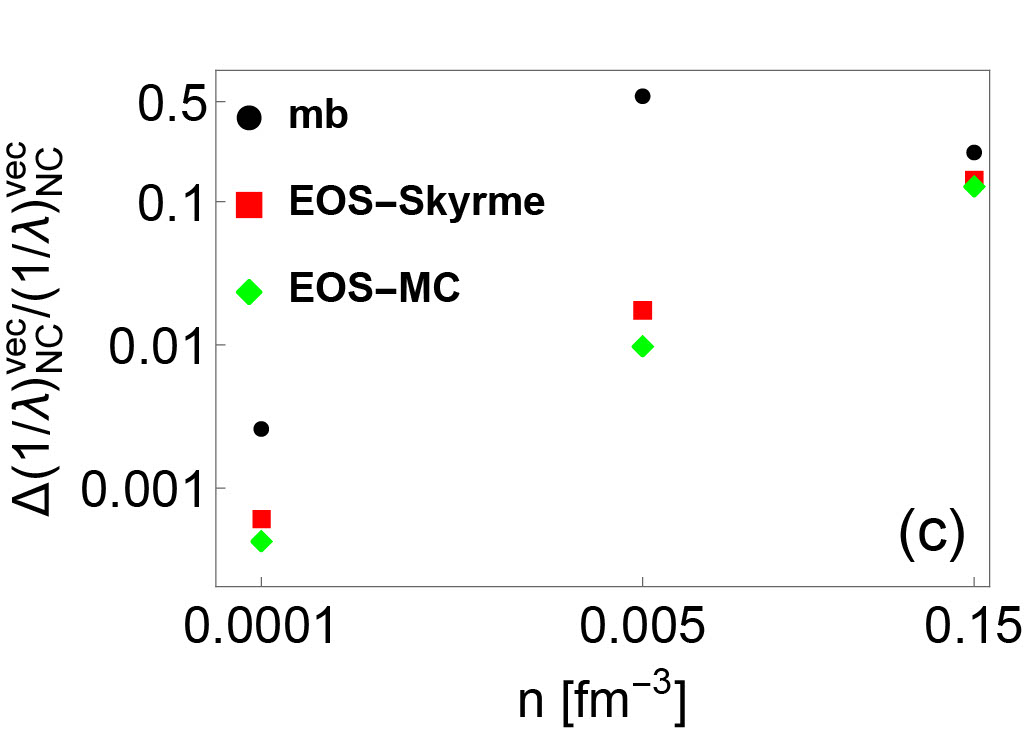}
	\includegraphics[width=0.45\textwidth]{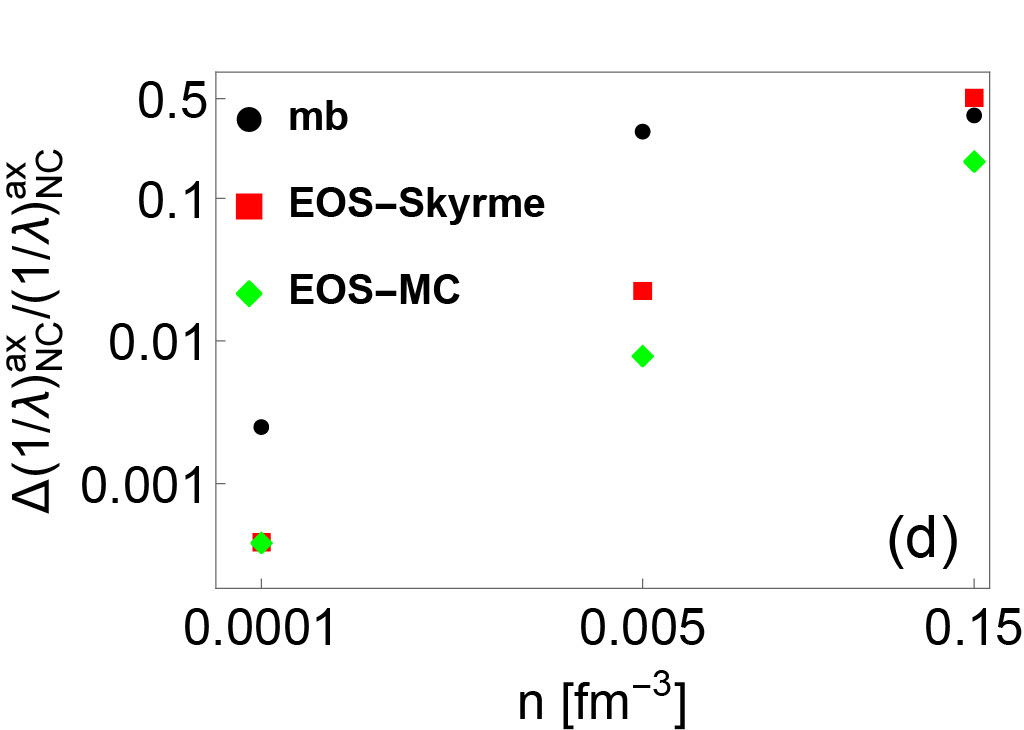}
	\caption{Differences in the neutrino inverse mean free path for charged- and neutral-current interactions, separated into vector and axial-vector components \citep{LiStMa23}, given variations in the equation of state resulting from variations in the nucleon--nucleon interaction adopted (EOS-Skyrme) or given variations in how the neutrino opacities are calculated (mb). The points labeled ``mb'' correspond to the magnitude of the difference between calculations performed using the Hartree--Fock (HF) approximation only, to correct for the nucleon effective mass resulting from nucleon--nucleon interactions, and the calculations performed correcting for both the nucleon effective mass (HF) and the effective weak interactions between neutrinos and nuclear matter, in the latter case using the Random Phase Approximation (RPA)---i.e., the calculations performed using HF+RPA.} 
\label{fig:imfpuncertainty}	
\end{figure*}

\subsection{Progenitors}

The lion's share of multidimensional core collapse supernova simulations, which have brought us to our current understanding of such supernovae, have been conducted using spherically symmetric progenitors. Given the obvious challenge of performing single-star simulations of stellar evolution in three dimensions and the even greater challenge of capturing the evolution in binary systems, it should not be surprising that efforts to capture the characteristics of three-dimensional progenitors, particularly as they pertain to the explosion mechanism, have proceeded with the more modest, but realizable, goal of modeling late-stage stellar evolution in three dimensions. Specifically, three-dimensional late-stage stellar evolution simulations have captured the dynamics of silicon- and oxygen-shell burning, resulting in turbulent convection, convective overshoot, and the mixing of nuclear species \citep{MuViHe16,FiCo21,YoTaKo21}. Such convection introduces large-scale, large amplitude asymmetries in the progenitor that have been demonstrated to alter the explosion dynamics not just quantitatively, but qualitatively \citep{MuMeHe17}. As discussed by \citet{FiCo20}, these asymmetries enhance the convection and the turbulence in the neutrino heating region below the shock, thereby increasing the total (thermal plus turbulent) stress acting on the shock. 

As late-stage stellar evolution is only part of the story, studies of this kind must continue and must extend in the same spirit to earlier stages, which in the end provide the initial conditions for late stages. The importance of three-dimensional stellar evolution has been demonstrated. What remains is an improved quantitative understanding of the initial conditions we should expect at the onset of core collapse. Most difficult here will be the determination of the initial differential rotation of the stellar core prior to collapse, as well as its initial magnetic field strength and topology, both essential initial conditions for the exploration of the roles of rotation and magnetic fields, including rare cases requiring initial rapid rotation and large-amplitude fields, in core collapse supernovae.

\section{Microphysics Input}

While developments in the microphysics included in core collapse supernova simulations is not the purview of the core collapse supernova modeling community and relies on progress made by the neutrino and nuclear physics communities, we would be remiss if a discussion of challenges and needs were not included here. 

\begin{figure}[h]
\centering
\includegraphics[width=0.48\textwidth]{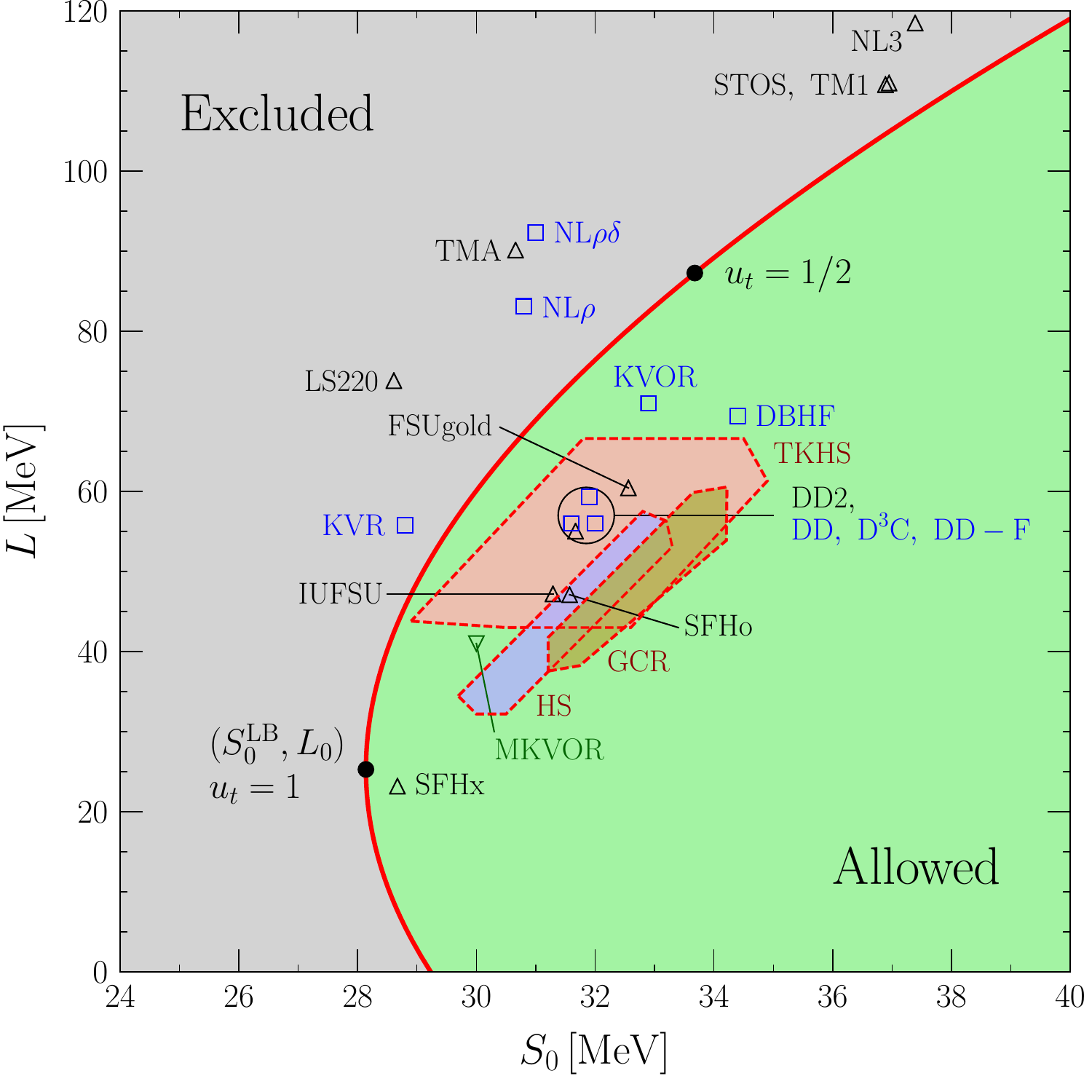}
\caption{Plot of the nuclear symmetry energy parameters $S_{0}$ and $L$ (capturing the change in the energy of nuclear matter as we move away from symmetric nuclear matter, which comprises equal numbers of neutrons and protons) for a number of nuclear equations of state, a subset of which have been implemented in core collapse supernova simulations. The gray shaded region is excluded by terrestrial experiment and astronomical observation. The gray shaded region contains a number of equations of state that had been used in past simulations---e.g., the widely used equation of state LS220. The green shaded region is allowed and contains several equations of state currently used in core collapse supernova simulation---e.g., DD2, FSUgold, IUFSU, SFHo, and SFHx.}
\label{fig:EOS}
\end{figure}

\subsection{Neutrino Weak Interactions}

For much of the first fifty of sixty years since the first simulations of core collapse supernovae were performed, neutrino weak interactions involving nucleons did not take into account interactions and correlations among them. \citet{RePrLa98} provided the first steps in this direction by modernizing the \citet{Bruenn1985} neutrino--nucleon charged-current absorption opacities and neutral-current scattering opacities, to include correlations for nucleon degeneracy (i.e., Fermi--Dirac statistics, the lowest-order correlations) and mean-field (MF) corrections resulting from nucleon--nucleon interactions, which are captured by a correction to the nucleon mass---i.e., an in-medium effective mass. The latter were calculated using the Hartree--Fock approximation. See Figure \ref{fig:HF} for the associated Feynman diagrams. These corrections were later implemented in core collapse supernova models by \citet{MuJaMa12} and shown to impact neutrino shock reheating. The electron neutrino-- and anti-neutrino--spheres are heated by the small-energy scattering of heavy-flavor neutrinos and antineutrinos on nucleons there resulting from these corrections, thereby boosting the electron-flavor neutrino luminosities and hardening their spectra, in turn resulting in increased neutrino shock reheating. \citet{RePrLa99} later included corrections to the weak interaction {\em per se} between the neutrinos and the now ``dressed'' nucleons (i.e., the nucleons now had an effective mass, not a bare mass) using the random phase approximation (RPA). Independently, \citet{BuSa98} modernized the rates for neutrino--nucleon neutral-current scattering including both MF and RPA corrections. \citet{RePrLa99} and \citet{BuSa99} also considered corrections for charged-current neutrino--nucleon absorption.

\begin{figure*}[!ht]
\includegraphics[width=0.5\textwidth]{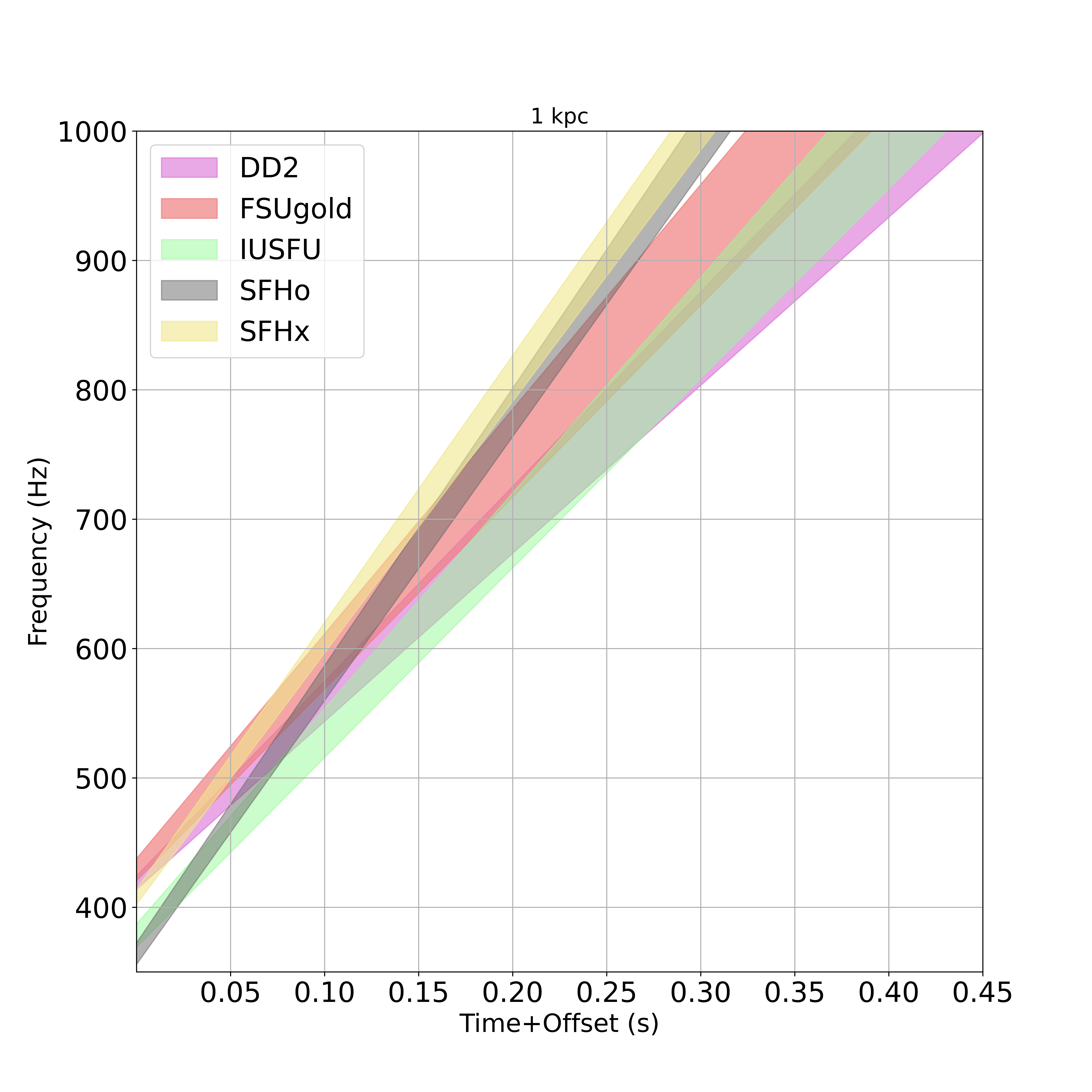}
\includegraphics[width=0.5\textwidth]{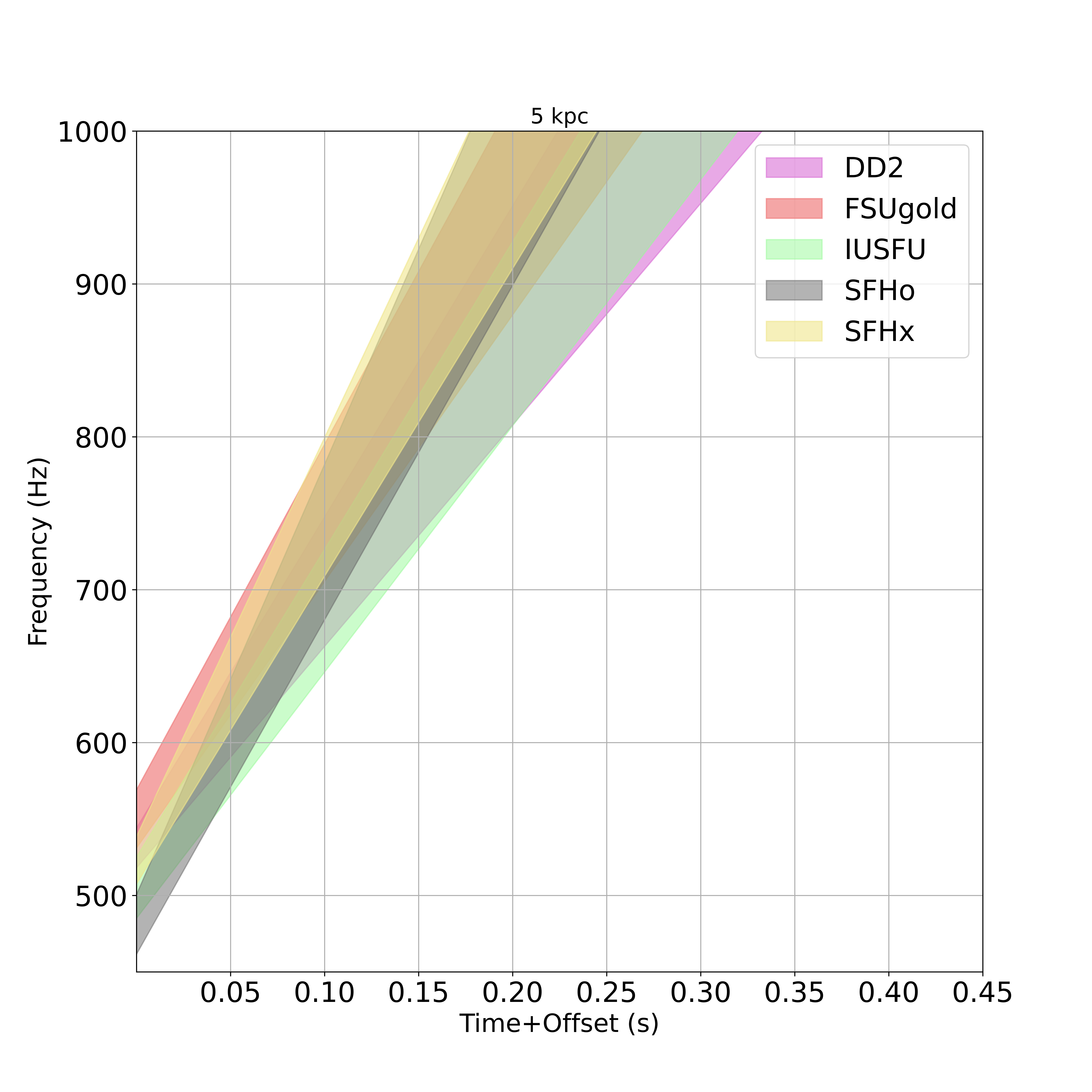}
\includegraphics[width=0.5\textwidth]{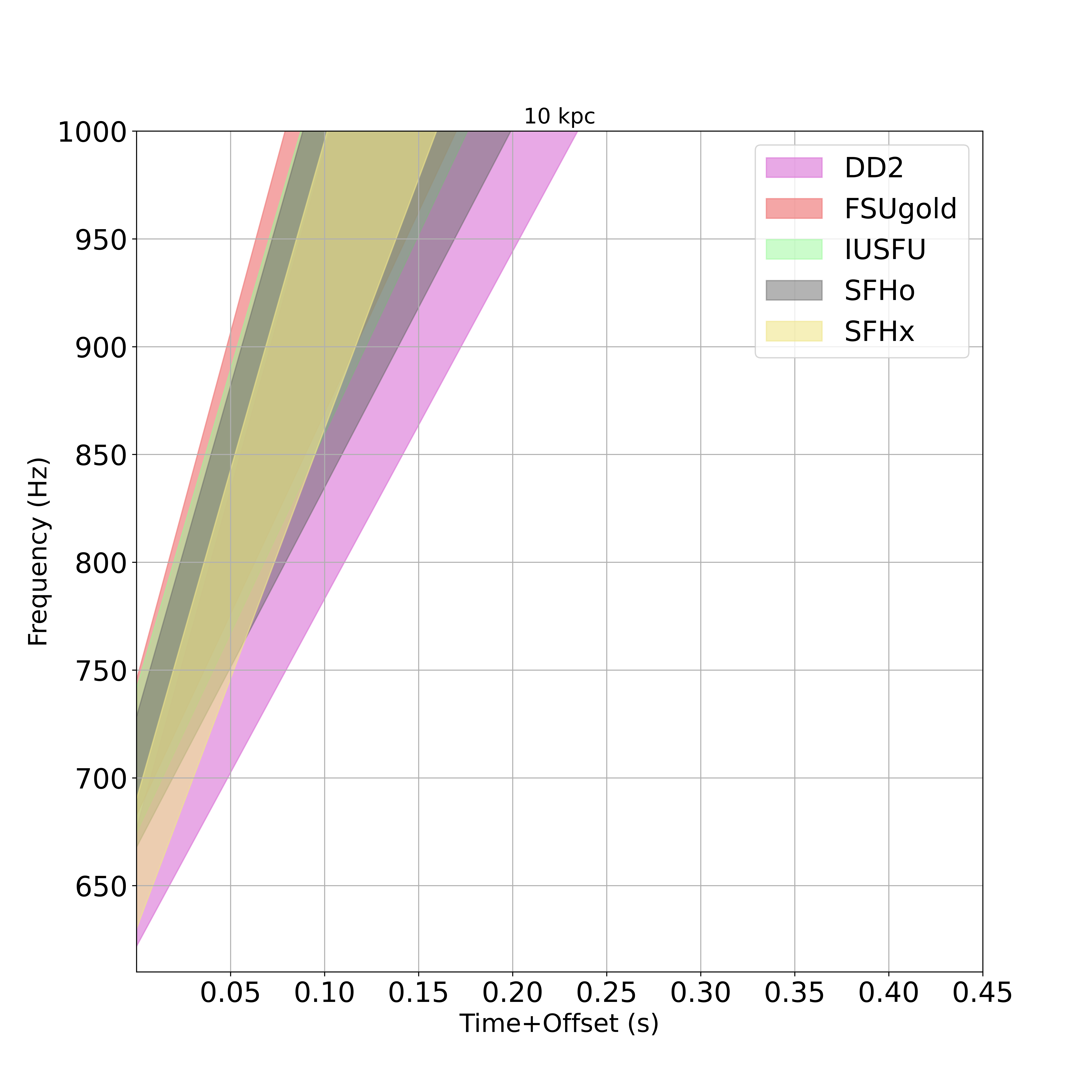}
\caption{Regions in the frequency--time domain corresponding to different slopes of the high-frequency, $gf$-feature in a core collapse supernova gravitational wave spectrogram, and their uncertainties, for different equations of state, at three different distances, taken from \citet{MuCaMe24}. The results at 1 kpc, which are expected for 10 kpc in next-generation detectors, indicate that discernment of equations of state is in principle possible.}
\label{fig:HFF_est_error}
\end{figure*}

The first important corrections to the \citet{Bruenn1985} rates for charged-current electron capture on nuclei, taking into account interactions among the nucleons, were developed by \citet{LaMaSa03} and in turn implemented in simulations of stellar core collapse by \citet{HiMeMe03}. These authors demonstrated that such correlations had a significant impact on the size of the inner, subsonically infalling core at bounce, impacting the shock formation radius and energy imparted to it---i.e., the initial conditions for the post-bounce evolution. As in the case of neutral-current neutrino--nucleon scattering, the electron capture rates in this case were obtained using RPA.

The RPA-associated Feynman diagrams are shown in Figure \ref{fig:ringdiagramsummation}. They are known as the ring diagrams. As in the case of Hartree--Fock corrections, which are included by summing to all orders in bubble and oyster diagrams, RPA corrections are included by (i) summing to all orders in ring diagrams and then (ii) replacing the interaction lines in the Feynman diagram on the left in Figure \ref{fig:electronneutrinoabsorption} by the effective interaction defined by Figure \ref{fig:ringdiagramsummation}. In RPA, the neutrino induces particle--hole pairs in the medium, which can induce other particle--hole pairs as the result of particle--hole interactions, as shown in Figure \ref{fig:ringdiagramsummation}, inducing long-range interactions and correlations. (Holes are defined relative to the nucleon Fermi sea.) RPA is the leading-order, self-consistent approximation to the response of MF-described nuclear matter to neutrino absorption or scattering.

Figure \ref{fig:imfpuncertainty} from \citet{LiStMa23} plots the differences in the neutrino inverse mean free path for both charged- and neutral-current interactions between calculations performed in MF and those performed in MF+RPA (points labeled ``mb'' for many-body effects) for several choices of the density. We can see that the differences are significant across charged- and neutral-currents, across vector and axial-vector couplings, and across densities. The differences are largest at higher densities, and at a given density they are significantly larger for charged-current interactions relative to neutral-current interactions. To date, RPA corrections have been included by some groups in neutral-current neutrino--nucleon scattering rates (\cite{BuSa98,Janka_2012}; see also the NuLib opacity library) but not in charged-current neutrino absorption.

Regarding corrections to charged-current neutrino interactions, further discussion is warranted. For illustrative purposes, consider the single-particle dispersion relation in the nonrelativistic MF approximation

\begin{equation}
E=\frac{\vec{p}^{2}}{2M^{*}}+U_{\rm p,n},
\label{eq:singleparticledispersion}
\end{equation}
where $M^{*}$ is the nucleon effective mass and $U_{\rm p,n}$ is the MF potential felt by a proton or neutron, respectively. The potential difference

\begin{equation}
dU\equiv U_{\rm n}-U_{\rm p}
\label{eq:dU}
\end{equation}
plays an important role in charged-current neutrino absorption \citep{RoReSh12} and is proportional to the not-well-constrained \citep{RrHoBa15} slope, $L$, of the symmetry energy of nuclear matter as a function of deviations from symmetric nuclear matter (i.e., equal numbers of protons and neutrons). Thus, the uncertainties associated with charged-current neutrino interactions are due to the uncertainties in both $dU$ (at the MF level) and the nucleon--nucleon interaction (at the MF+RPA level).

Given the above discussion, two things should be emphasized:
\begin{enumerate}
    \item To date, RPA corrections have not been fully implemented (i.e., for both charged and neutral currents) in neutrino opacity tables used in core collapse supernova simulations.
    \item Given that RPA corrections are sensitive to the poorly constrained spin-dependent part of the nucleon--nucleon interaction, as well as to the poorly constrained MF potential difference between protons and neutrons, they themselves are poorly constrained. [See the discussions in \citet{RePrLa99} and \citet{LiStMa23} and in \citet{LiCoSt26} a discussion of recent efforts to better constrain the spin-dependent part of the nucleon--nucleon interaction.]
\end{enumerate}
Regarding the first point, the studies thus far suggest that RPA corrections are important at high densities, where neutrino diffusion time scales initially decouple the impact of the corrections from neutrino emissions at the neutrinospheres for early postbounce times. However, multidimensional modeling has shown that explosions and, more specifically, explosion energies, develop over several seconds, during which time RPA corrections will likely impact neutrino emissions and, in turn, shock heating. Taken together, the above points make clear there is still a lot of room for improvement with regard to pinning down the neutrino opacities of relevance for core collapse supernovae and that we might anticipate nontrivial changes in the models in the future.

The neutrino opacities and the nuclear equation of state are intimately related. Calculations of MF and RPA corrections to the opacities necessitate a choice of nucleon--nucleon interaction, which is foundational to constructing the energy per baryon of nuclear matter, upon which any equation of state is built. A discussion of the underlying uncertainties in calculations of the neutrino opacities without a discussion of the underlying uncertainties in the determination of the nuclear equation of state would be incomplete and lack self consistency. The uncertainties in both are rooted in the same thing, uncertainties in the nucleon--nucleon interaction. This is no better illustrated than in Figure \ref{fig:imfpuncertainty}. Points labeled EOS-Skyrme correspond to differences in the inverse neutrino mean free path resulting from different choices of the nucleon--nucleon interaction (in this case, Skyrme interaction), as a function of density, weak currents, and vector- and axial-vector couplings. At higher densities, these differences/uncertainties become comparable to the differences/uncertainties associated with many-body effects.

\subsection{The Nuclear Equation of State}

Terrestrial experiment and astronomical observation continue to constrain the nuclear equation of state of neutron-rich, dense matter in core collapse supernovae and in neutron stars and their mergers. Equations of state that have been ruled out by virtue of low- and high-energy nuclear experiment and astronomical observations of neutron star masses and radii are shown in the gray shaded region of Figure \ref{fig:EOS}, from \citet{TeLaOh17}. The green shaded region, on the other hand, contains nuclear equations of state that, to date, are admissible. To highlight the dependence of the nuclear equation of state and, in turn, the neutrino opacities on the treatment of the nucleon--nucleon interaction, we focus on the admissible equations of state that derive from the equation of state of Hempel--Schaffner-Bielich \citep{HeSc10}: DD2, FSUgold, IUFSU, SFHo, and SFHx. They are widely used equations of state in core collapse supernova simulations and differ primarily by the parameterization used in their relativistic mean field treatments of nucleons (the names of the five equations of state correspond to the names of the five different parameterizations of the nucleon--nucleon interaction used). Notably, in the cases of SFHo and SFHx, the parameterizations are based on observations of neutron star masses and radii, differentiating these two equations of state from the rest, where the parameterizations are based terrestrial experiment.

In the cases considered here, the five different parameterizations reflect both uncertainty and opportunity. The nucleon--nucleon interaction is overall not very well constrained, though it is better constrained in the spin-independent channel than in the spin-dependent channel \citep{LiStMa23}. Advances on this front will take time, and core collapse supernova simulations will have to be performed with all admissible equations of state to assess the impact of the uncertainty in the equation of state on supernova outcomes. On the other hand, given the developing sophistication of core collapse supernova models, and given the subset of equations of state discussed here, which differ only in the choice of the nucleon--nucleon interaction, it may be possible to glean something of the nucleon--nucleon interaction given a Galactic supernova and the associated detection of neutrino ``light curves'' of all three flavors and of the gravitational waves from both matter and neutrinos, though many factors will contribute to both the neutrino and the gravitational wave emission of the event. An example of this is presented in \citet{MuCaMe24}, where in the context of a series of two-dimensional core collapse supernova simulations that implemented six different equations of state while holding everything else fixed it was shown that the slope of the ubiquitous $gf$-feature in the gravitational wave spectrograms depends on the equation of state adopted. It was further shown that in next-generation detectors differences in the slopes associated with different equations of state would be discernible, thereby providing, in principle, a means to discern between equations of state. (In practice, the slope depends on multiple factors that would need to be disentangled using other observations before statements about the equation of state could be made.) 

\section{Outlook}

Exponential progress over the past decade has advanced core collapse supernova theory considerably. Sophisticated simulations in three spatial dimensions have provided a great tool to decipher the explosion mechanism and to produce detailed predictions for neutrino and gravitational wave emission. On the other hand, this progress was made in part through approximations that enabled tools from spherical symmetry to be deployed in multidimensional simulations. Such tools were painstakingly developed over decades, reflecting the level of difficulty of this problem. Eventually, these approximations, and others, must be lifted. Doing so defines a well-illuminated path forward over the next 5--10 years, leading to three-dimensional, general relativistic spectral two-moment neutrino radiation hydrodynamics (or magnetohydrodynamics) simulations of core collapse supernovae from all supernova groups.

While such simulations would mark another significant milestone in core collapse supernova theory, they would not mark the end of the developments needed. Two-moment models are limited by the closure problem. The two-moment approach must ultimately give way to multi-angle, spectral Boltzmann kinetics. However, the implementation of Boltzmann kinetics throughout a core collapse supernova simulation must wait for supercomputing architectures able to render such an implementation practical. 

Looking even further down the road, Boltzmann kinetics would provide an ideal foundation to continue to advance the study of neutrino flavor transformation in core collapse supernova environments. A great deal has been learned from current implementations, some based on moments methods. Of greatest importance, these efforts have demonstrated that further investigation of the impact of flavor mixing, particularly fast flavor mixing, on neutrino shock reheating is necessary. Current implementations must give way to full quantum kinetics implementations describing the evolution of the neutrino density matrix of distribution functions. This is a daunting challenge, one that may be met in the foreseeable future only through the use of subgrid models given the length and time scales over which fast flavor transformation may occur, both orders of magnitude smaller than what can currently be resolved. 

Fortunately, such increasing challenges will be met with numerical methods of increasing sophistication, with computing platforms of increasing capability, with the core collapse supernova modeling community's ever increasing experience, and with the increasing knowledge gained of core collapse supernova dynamics as such challenges are taken on, by design, in achievable stages.

Last but not least, this multidimensional challenge will clearly not be met by the core collapse supernova modeling community alone. Our success to date is in no small part due to the efforts of the experimental and theoretical low- and high-energy nuclear physics communities. To date, the nuclear equation of state has been constrained significantly and the calculations of neutrino opacities in interacting and correlated nuclear matter have advanced significantly through these efforts. Here, too, long-term challenges remain, largely rooted in the long-term challenge of further constraining the nucleon--nucleon interaction, especially its spin dependence. All communities involved will ultimately be, scientifically speaking, at the receiving end of a once-in-a-generation payoff when we are graced with the next Galactic core collapse supernova and supernova, nuclear, and neutrino physics will be tested like never before.

\section*{Acknowledgments}

AM was supported in part by grants from the National Science Foundation Gravitational Physics Theory Program through awards PHY-1806692, PHY-2110177, and PHY-2409148. The author would like to thank Lucas Johns, Zidu Lin, Bernhard Mueller, Gerard Nav\'{o}, and Ingo Tews for permission to include their figures here. The author would especially like to thank Zidu Lin for extensive discussions regarding neutrino opacities in nuclear matter.

\end{document}